\documentclass[conference]{IEEEtran}

\usepackage{amsmath,amsthm,amssymb}
\usepackage{graphicx}

\newtheorem{definition}{Definition}

\usepackage{multirow}
\usepackage{tabularx}
\usepackage[ruled,vlined]{algorithm2e}

 \makeatletter
 \let\@copyrightspace\relax
 \makeatother

\begin{document}

\title{Low-Complexity Cloud Image Privacy Protection via Matrix Perturbation}

\author{\IEEEauthorblockN{
Xuangou~Wu\IEEEauthorrefmark{1},
Shaojie~Tang\IEEEauthorrefmark{2},
and Panlong~Yang\IEEEauthorrefmark{3},}
\IEEEauthorblockA{\IEEEauthorrefmark{1}the School of Computer Science and Technology, AHUT, Ma'anshan, China}
\IEEEauthorblockA{\IEEEauthorrefmark{2}University of Texas at Dallas, USA}
\IEEEauthorblockA{\IEEEauthorrefmark{3}the Institute of Communication Engineering, PLAUST, Nanjing, China}
\IEEEauthorblockA{Emails: wxgou@mail.ustc.edu.cn, tangshaojie@gmail.com, panlongyang@gmail.com}
}

\maketitle

\begin{abstract}
Cloud-assisted image services are widely used for various applications. Due to the high computational complexity of existing image encryption technology, it is extremely challenging to provide privacy preserving image services for resource-constrained smart device. In this paper, we propose a novel encrypressive cloud-assisted image service scheme, called eCIS.  The key idea of eCIS is to shift the high computational cost to the cloud allowing reduction in complexity of encoder and decoder on resource-constrained device. This is done via compressive sensing (CS) techniques, compared with existing approaches, we are able to achieve privacy protection at no additional transmission cost. In particular, we design an encryption matrix  by taking care of  image compression and encryption simultaneously. Such that, the goal of our design is to minimize the mutual information of original image and encrypted image. In addition to the theoretical analysis that demonstrates the security properties and complexity of our system, we also conduct extensive experiment to evaluate its performance. The experiment results show that eCIS  can effectively  protect image privacy and meet the user's adaptive secure demand. eCIS reduced the system overheads by up to  $4.1\times\sim6.8\times$ compared with the existing CS based image processing approach.
\end{abstract}


\section{Introduction}
During the recent years, cloud-assisted image services are widely used for various applications, such as individual image storage and sharing by smartphone (e.g., Facebook and QQ),  healthcare image monitoring by wireless sensors \cite{shoaib2011digital, alemdar2010wireless} and so on. However, this also brings serious security threat to users because of the public access of cloud \cite{armbrust2010view}. Meanwhile, more and more resource-constrained smart devices are used as sensors for sampling these image signals, this also poses a great challenge to the development of appropriate image encryption techniques for mobile device.

Existing cloud security image encryption techniques usually follow compression then encryption paradigm. These techniques require image acquisition device to perform either expensive coding or encryption operation \cite{liu2011image,wang2011new} on original image.
For example,  transformed-based image coding techniques require complex encoding computation.
In recent years, compressive sensing (CS) has bee proposed for sparse/compressible signal sampling and compression.  C. Wang \emph{et.al.} proposed a cloud-assisted computation outsourcing scheme for healthcare video monitoring \cite{wang2014privacy}.  They implemented image encryption by  linear programming (LP) problem transformation, which is the application of LP secure outsourcing problem \cite{wang2011secure}. However, the transformation overhead may easily outweigh the benefits brought by their outsourcing scheme. 

As a result, existing results are not appropriate for resource constrained smart device as it requires low-complexity image compression and encryption. In particular, we notice that existing techniques require expensive computation resource mainly due to the following two separated processes: image compression and encryption. To this end, we propose a novel \textbf{e}ncrypressive \textbf{C}loud-assisted \textbf{I}mage \textbf{S}ervice scheme (eCIS). In eCIS, we implement image compression and encryption simultaneously via CS, which can significantly reduce the resource consumption for sampling and receiver devices. Meanwhile, we implement cloud storage and computation outsourcing for image user at no additional communication cost. The development of eCIS faces several unique challenges:
(1) How to design encryption matrix  to meet  CS theory and image signal encryption  without increasing transmission cost?
(2) How to implement encryption matrix to meet low-complexity requirement for encoder and end user?
(3) How to guarantee user's privacy performance with our design encryption matrix?

To address the above challenges, we design an encryption matrix based on inverse matrix and mutual information to meet CS encryption encoding and decoding at no additional transmission cost. To achieve low-complexity encryption and decryption for encoder and end user, we exploit adaptive perturbation matrix as our encryption matrix. Moreover, we conduct theoretical analysis that demonstrates several   security properties and low-complexity of our scheme.
The contributions of this paper are as the follows.

1) We present a novel encrypressive cloud-assisted image service scheme  via CS, called eCIS. Our scheme implements the image signal storage and computation outsourcing for both sampling device and end user. Meanwhile, it also can protect user's image signal privacy.

2) We formulate our security problem according to security and system requirements, and design an encryption matrix based on mutual information. Our designed encryption matrix implements compression and encryption simultaneously. We also conduct theoretical analyses of both image security and system overhead.

3) With the extensive experiments, we show that  eCIS not only protect image privacy efficiently but also meet the user's adaptive security demand. The experimental results also display that  eCIS decrease $4.1\times\sim6.8\times$ time cost for sampling device and end user compared with the existing CS based image processing approach.

The rest of this paper is organized as follows.  In section 2 presents the related preliminaries. The problem statement is given in section 3. The detailed design of eCIS is presented in Section 4. In section 5, we give the theoretical performance analysis of eCIS. Section 6 reports our experimental results. We present a literature review of existing work in section 7. Finally, we make a conclusion and future work in section 8.

\section{Preliminaries}
\subsection{CS and Compressive Signal}
CS theory asserts that a relatively small number linear combination of a sparse signal could contain most of its salient information \cite{donoho2006compressed}. This technique shifts the computation cost from encoder to decoder compared with transformed-based image compression \cite{baraniuk2007compressive}.
Assuming that $\mathbf{s}\in\mathbb{R}^n$ is a $t$-sparse signal, which has only  $t$ non-zero components. Thus, the information can be extracted from $\mathbf{s}$ by
$
\mathbf{y}=\Phi\mathbf{s}
$,
where $\Phi$ is an $m\times n$ measurement matrix, $\mathbf{y}\in\mathbb{R}^m$ is measurement vector and $m\ll n$.  If $\Phi$ satisfies the restricted isometric property (RIP) and {\small $m\geq O(t\cdot log(n/t))$ } \cite{candes2005decoding}, the sparse signal $\mathbf{s}$ could be recovered with high probability.
Cand$\grave{e}$s, Romberg, and Tao \cite{candes2006robust} and Donoho \cite{donoho2006compressed} have shown many random matrices that satisfy the RIP such as Gaussian identity distribution matrix, $\pm$1 Bernoulli matrix and so on. The signal $\mathbf{s}$ could be recovered via $\ell_1$ optimization as
\begin{equation*}
\hat{\mathbf{s}}=\underset{\mathbf{s}}{\arg\min}\parallel\mathbf{s}\parallel_{1}\,\,\, s.t.\,\,\,\mathbf{y}=\Phi\mathbf{s}
\end{equation*}
There have been many efficient algorithms to solve the above problems such as basis pursuit \cite{candes2005decoding}, orthogonal matching pursuit (OMP) algorithm \cite{OMP}, CoSaMP \cite{needell2009cosamp} and so on.

The real image signal, however, is rarely  sparse, which can be transformed into sparse signal by a sparse representation basis.  In other words, this signal can be well-approximated by a sparse signal, which is called compressible signal. For example, an image signal $\mathbf{x}\in \mathbb{R}^n$ can usually be transformed into a sparse signal $\mathbf{s}$ under  discrete cosine transformation (DCT) basis or discrete wavelet transformation (DWT) basis. Given $\mathbf{x}=\Psi\mathbf{s}$ and $\Psi$ is a $n\times n$ representation basis, $\mathbf{s}=[s_1,s_2,\cdots,s_n]^T$ is the coefficient vector of $\mathbf{x}$ under $\Psi$. If $\mathbf{x}$ is compressible, then the magnitudes of the sorted coefficients $s_i$ observe a power-law decay :
 \begin{equation*}
    |s_i|\leq C\cdot {i}^{-q}
 \end{equation*}
where $C$ and $q>0$ are constants. 
The compressible signal $\mathbf{x}$ can be represented accurately by only $t$ ($t\ll n$) coefficients \cite{compsig1998}.

\subsection{Mutual Information}
In information theory, mutual information represents the mean relevance of two random variables which is defined as follows.
\begin{definition} \label{def_mu} \cite{elementIt2006}
Consider two random variables $X$ and $Y$ with a joint probability mass function $p(xy)$, marginal probability mass functions $p(y)$ and conditional probability function $p(y/x)$. The mutual information $I(X; Y)$ is :
\begin{eqnarray}\label{e_mut_def}
I(X; Y) & = & \underset{x\in X}{\sum}\underset{y\in Y}{\sum}p(xy)\log_{2}\frac{p(y/x)}{p(y)}\nonumber \\
 & = & -\underset{y\in Y}{\sum}p(y)\log_{2}p(y)+\underset{x\in X}{\sum}\underset{y\in Y}{\sum}p(xy)\log_{2}p(y/x)\nonumber \\
 & = & H(Y)-H(Y/X)
\end{eqnarray}
where $H(Y)$ and $H(Y/X)$ represent entropy and conditional entropy, respectively.
\end{definition}

The smaller $I(X;Y)$ is, the less the information of $X$ obtained from $Y$. If $X$ and $Y$ are two  independent random variables, $I(X;Y)$ equals to zero. It means that no information of $X$ could be obtained from $Y$.

\begin{figure}
\begin{center}
  \includegraphics[scale=0.6]{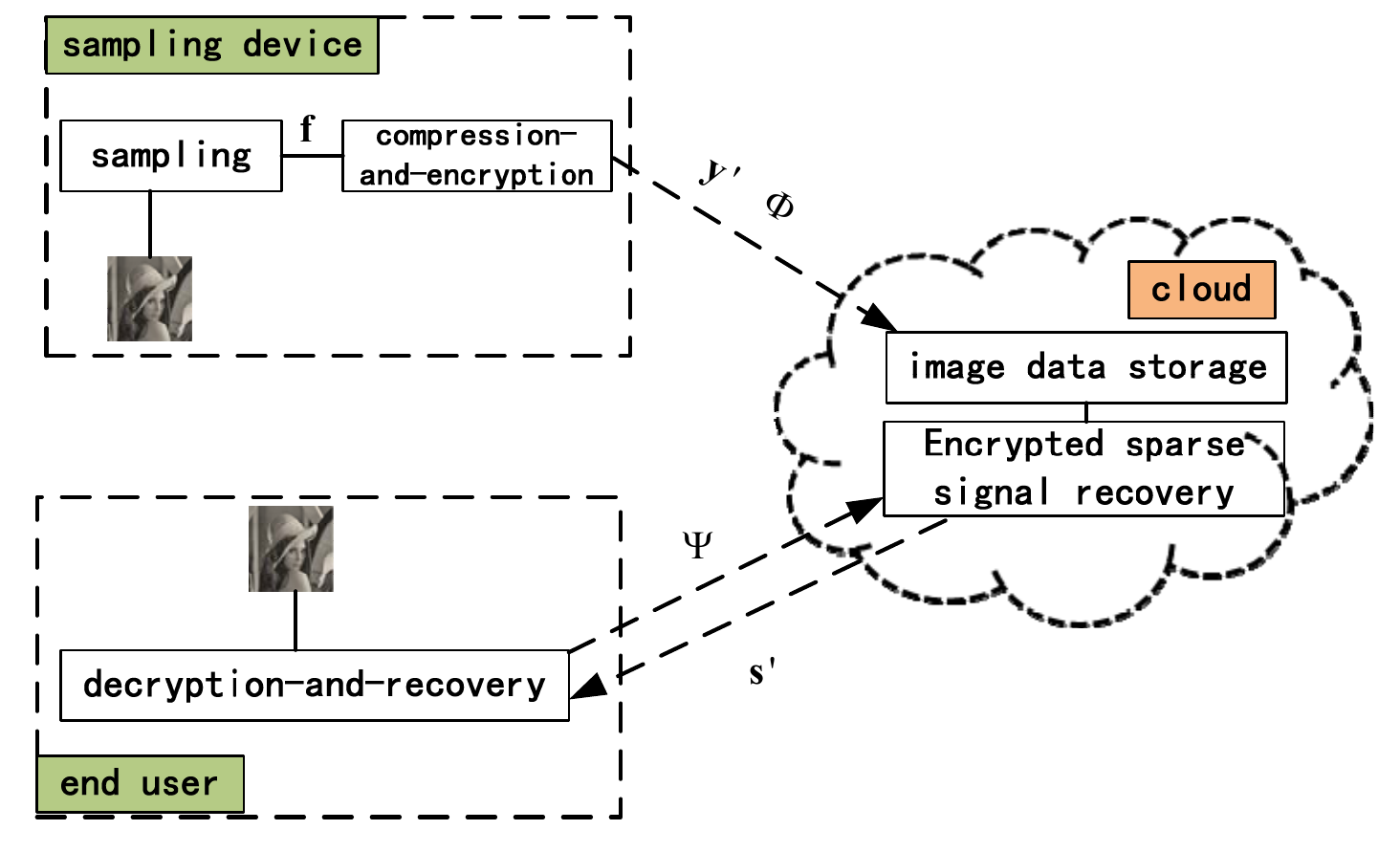}\\
  \caption{System architecture of  eCIS.}\label{f_arc}
\end{center}
\end{figure}

\section{Problem statement}
\subsection{System Architecture}
Fig.\ref{f_arc} shows the overall architecture of eCIS. The system consists of three parties: sampling device, cloud, and end user.
The sampling device is responsible for sampling, compressing, and encrypting the original image signal. After image sampling, the device carries out compression and encryption operations.  Then the sampling device uploads its compressed and encrypted signal $\mathbf{y}'$ and  measurement matrix $\Phi$ to the cloud. If an end user wants to access the image, he should send its request and sparse representation basis, $\Psi$, to the cloud. Cloud receives the user's image request and $\Psi$, and calculates the sparse signal $\mathbf{s}'$ via CS decoding algorithm and sends it to the end user.
After receiving $\mathbf{s}'$, the end user is responsible for recovering the original image $\mathbf{f}$  by $\mathbf{s}'$.
We assume that the key of encryption matrix cloud be securely transmitted between the sampling device and end user. In practice,  sampling device and end user could be the same smart device(e.g., smartphone).

\textbf{Design goals:}
Our goals  consist of three aspects:
(1) The sampling device should implement image compression and protect the user's image privacy. Meanwhile, the compression and encryption process should have low complexity and transmission cost.
(2) The cloud  should carry out complicated CS-based image decoding without leaking user's privacy.
(3) The image recovery and decryption at end user should also have low complexity.

\subsection{CS-based Image Compression}
Before the problem formulation, we first introduce several important notations used in CS-based image compression. The original image signal and its sparse representation basis are denoted by $\mathbf{f}$ and $\Psi$, respectively. The corresponding sparse signal of $\mathbf{f}$ under $\Psi$ is denoted by $\mathbf{s}$, namely, $\mathbf{s}=\Psi^{-1}\mathbf{f}$.

Firstly, we review CS-based image compression and decoding for image signal $\mathbf{f}$. The compression process is expressed as
\begin{equation}\label{e_cs_enc}
    \mathbf{y}=\Phi \mathbf{f}=\Phi\Psi \mathbf{s}
\end{equation}
where $\mathbf{y}$ presents the measurement vector. The decoding process is expressed as
\begin{equation}\label{e_cs_dec}
   \hat{\mathbf{s}}=\underset{\mathbf{s}\in\mathbb{R}^n}{\arg\min}\|\mathbf{s}\|_1 ~~~~s.t. ~~~~\mathbf{y}=\Phi\Psi \mathbf{s}
\end{equation}
where $\hat{\mathbf{s}}$ is the recovery sparse signal corresponding to $\mathbf{f}$, with $\hat{\mathbf{f}}=\Psi\hat{\mathbf{s}}$.
If the attacker does not know  $\Phi$ and  $\Psi$, the measurement vector $\mathbf{y}$ can be considered as ciphertext  \cite{orsdemir2008security}. 
However, if the recovery process is carried out in cloud, the measurement vector $\mathbf{y}$ is no longer private  because both  $\Phi$ and $\Psi$ are public.
Then how to outsource expensive CS decoding task to the cloud without revealing the original image signal? Since the measurement matrix has a strong encryption function, can we consider different measurement matrices for encoding and decoding of sampling device and cloud?

Assuming that the decoding measurement matrix used in cloud is $\Phi$, which could be general measurement matrix such as Gaussian random matrix, $\pm1$ Bernoulli matrix \cite{candes2006robust}. The encoding measurement matrix is denoted by $\Phi A$, where $A$ is a $n\times n$ matrix. Then, the image compression process is expressed as
\begin{equation}\label{e_cry_cs_enc1}
    \mathbf{y}'=\Phi A f
\end{equation}
where $\mathbf{y}'$ is measurement vector corresponding to $\Phi A$. Due to $\mathbf{f}=\Psi \mathbf{s}$, Eq. \ref{e_cry_cs_enc1} can be expressed as
\begin{equation}\label{e_cry_cs_enc2}
    \mathbf{y}'=\Phi A \mathbf{f}=\Phi A \Psi \mathbf{s}
\end{equation}
If $A$ is invertible, $A\Psi$ can be converted as $\Psi B$, with $B=\Psi^{-1}A\Psi$.  Eq. \ref{e_cry_cs_enc2} is equivalent to
\begin{equation}\label{e_cry_cs_enc3}
    \mathbf{y}'=\Phi A \mathbf{f}=\Phi\Psi B\mathbf{s}=\Phi\Psi \mathbf{s}'
\end{equation}
where  $\mathbf{s}'=B\mathbf{s}$. After image compression, the cloud carries out the decoding operation as
\begin{equation}\label{e_cry_cs_dec}
   \hat{\mathbf{s}'}=\underset{\mathbf{s}\in\mathbb{R}^n}{\arg\min}\|\mathbf{s}'\|_1 ~~~~s.t. ~~~~\mathbf{y}'=\Phi\Psi \mathbf{s}'
\end{equation}
where $\hat{\mathbf{s}'}$ is the recovery signal corresponding to $\mathbf{s}'$ via solving $\ell_1$ optimization problem.
Therefore, if we can find a suitable matrix $A$ satisfied three conditions: (1) $A$ is invertible. (2) $A$ can not decrease the sparsity of $\mathbf{s}$. (3) $\mathbf{s}$ can not be obtained from $\mathbf{s}'$ without $A$. We can implement secure image data storage and CS decoding in the cloud.

\subsection{Problem Formulation}
We next introduce the problem formulation. In the rest of this paper, $\Phi$ and $A$ are called measurement matrix and encryption matrix, respectively. If an attacker can not recover $\mathbf{f}$ given $\mathbf{s}'$ and $\Psi$, we can securely implement image decoding in the cloud. Therefore, our problem is to find an appropriate encryption matrix $A^*$ such that
\begin{eqnarray*}\label{e_obj_fun}
\mathbf{(P)}~~~~  A^* & = &  \underset{A\in\mathbb{R}^{n\times n}}{\arg\min}~~~\mathcal{P}\left(A,\Phi,\mathbf{f}\right)  \\
   & s.t.& ~~~\mathbf{y}'=\Phi A\mathbf{f}=\Phi \Psi \mathbf{s}'  \\
   & &     ~~~\mathbf{s}'=\Psi^{-1}A\Psi \mathbf{s}  \\
   & &     ~~~\mathcal{C}(\Phi, A, \mathbf{f})\leq \mathcal{C}(\Phi, \mathbf{f})
\end{eqnarray*}
where $\mathcal{P}(\cdot)$ is privacy exposure function which will be described later, $\mathcal{C}(\cdot)$ is communication cost function. $\mathcal{C}(\Phi, A, \mathbf{f})$ $\leq  \mathcal{C}(\Phi, \mathbf{f})$ ensures  that the transmission cost of our scheme is no larger than the original CS based image compression. 

\section{Our Solution}
In this section, we will discuss our encryption matrix design and implement in detail, and give each component design of our eCIS.
\subsection{Encryption Matrix Design and Implement}
\subsubsection{Design Principle}
Recall that our goal of encryption matrix is to ensure that the original sparse signal of user's image can not be obtained from the recovered sparse signal in the cloud. If the encryption matrix does not change the sparsity of original image signal, we can remove the condition of $\mathcal{C}(\Phi, A, \mathbf{f})\leq \mathcal{C}(\Phi, \mathbf{f})$ from the problem (\textbf{P}). This is because when the measurement matrix is given, the number of CS measurements which  only depends on the sparsity of the compressed signal.

In information theory, mutual information represents the shared information between two random variables according to Definition \ref{def_mu}. In this work, we exploit mutual information as our privacy exposure function. Intuitively, it measures what extent $\mathbf{s}$ can be inferred from $\mathbf{s}'$. We assume that  $\mathbf{s}=[s_1,s_2,\cdots,s_n]^T$,   $\mathbf{s}'=[s'_1,s'_2,\cdots,s'_n]^T$,  $s_i\in\{a_1,a_2,$ $\cdots,a_m\}$, and $s'_i\in\{b_1,b_2,$ $\cdots,b_m\}$ for $i=1,2,\cdots,m$. each of the presentation, we also assume $\mathbf{\xi}_A=\{a_1,a_2,$ $\cdots,a_m\}$, $\mathbf{\xi}_B=\{b_1,b_2,$ $\cdots,b_m\}$, $P\{s_k=a_i\}=p(a_i)$,  and $P\{s'_k=b_j\}=p(b_j)$. 
Accordingly, the problem ($\mathbf{P}$) is equivalent to
\begin{eqnarray*}\label{e_mu_p}
    \mathbf{(P1)}~~~~   &  &      \underset{P(\mathbf{s}'/\mathbf{s})}{\min} I(\mathbf{s};\mathbf{s}')\\
   & s.t.& ~~~\mathbf{s}'=B \mathbf{s} \\
   & &    ~~~B=\Psi^{-1}A\Psi
\end{eqnarray*}
According to Definition \ref{def_mu}, $I(\mathbf{s};\mathbf{s}')=H(\mathbf{s}')-H(\mathbf{s}'/\mathbf{s})$. Since $\mathbf{s}$ and $\mathbf{s}'$ can be considered as discrete memoryless $n$ times extension of single symbol, $H(\mathbf{s}')$ is given by
\begin{eqnarray*}
    H(\mathbf{s}')&=&-\sum_{i_1=1}^{n}\cdots\sum_{i_n=1}^{n}p(a_{i_1}\cdots a_{i_n})\log p(a_{i_1}\cdots a_{i_n})  \\
    &=& -\sum_{i_1=1}^{n}\cdots\sum_{i_n=1}^{n}\prod_{k=1}^n p(a_{i_k})\log \prod_{k=1}^n p(a_{i_k})  \\
    &=& n H(\xi_B)
\end{eqnarray*}
Similarly, $H(\mathbf{s}'/\mathbf{s})$ is given by
{\small
\begin{eqnarray*}
  H(\mathbf{s}'/\mathbf{s}) &=& -\sum_{i_{1}=1}^{n}\cdots \sum_{i_{n}=1}^{n}\sum_{j_{1}=1}^{n}\cdots \sum_{j_{n}=1}^{n}p(a_{i_1}\cdots a_{i_n})  \\
          & & p(b_{j_1}\cdots b_{j_n}/a_{i_1}\cdots a_{i_n})\log_2p(b_{j_1}\cdots b_{j_n}/a_{i_1}\cdots a_{i_n})   \\
          &=& -\sum_{i_{1}=1}^{n}\cdots \sum_{i_{n}=1}^{n}\sum_{j_{1}=1}^{n}\cdots \sum_{j_{n}=1}^{n}p(a_{i_1}\cdots a_{i_n}) \\
          & &   \prod_{k=1}^n p(a_{i_k})p(b_{j_k}/a_{i_k}\log \prod_{k=1}^n p(b_{j_k}/a_{i_k})  \\
          &=& n(H(\mathbf{\xi}_B)-H(\mathbf{\xi}_B/\mathbf{\xi}_A))
\end{eqnarray*}}

According to Eq.\ref{e_mut_def}, $I(\mathbf{s};\mathbf{s}')$ is given by
\begin{eqnarray*}
    I(\mathbf{s};\mathbf{s}') & = & H(\mathbf{s}')-H(\mathbf{s}'/\mathbf{s}) \\
     & = & n(H(\mathbf{\xi}_B)-H(\mathbf{\xi}_B/\mathbf{\xi}_A))
\end{eqnarray*}
In order to minimize $I(\mathbf{s};\mathbf{s}')$, it is equivalent to minimizing $H(\mathbf{\xi}_B)-H(\mathbf{\xi}_B/\mathbf{\xi}_A)$. When $H(\mathbf{\xi}_B/\mathbf{\xi}_A)=H(\mathbf{\xi}_B)$, $I(\mathbf{s};\mathbf{s}')$ achieves the minimum value ($I(\mathbf{s},\mathbf{s}')=0$), this is because $H(\mathbf{\xi}_B)$ and $H(\mathbf{\xi}_B/\mathbf{\xi}_A)$ are greater than or equal to $0$. Note that $H(\mathbf{\xi}_B/\mathbf{\xi}_A)=H(\mathbf{\xi}_B)$ implies that  $p(b_j/a_i)=p(b_j)$ for $i,j=1,2,\cdots,n$.  Namely, the events $a_i$ and $b_j$ are independent to each other. In other words, $\mathbf{s}$ can not be recovered from $\mathbf{s}'$ if all elements  are independent to each other.

The goal of encryption matrix is to minimize $I(\mathbf{s};\mathbf{s}')$. When the image signal is given, the value of $I(\mathbf{s};\mathbf{s}')$ is decided by the conditional probability transform matrix, $\Gamma$, which is denoted by
\begin{eqnarray*}
\Gamma & = & \left[\begin{array}{cccc}
p(b_{1}/a_{1}) & p(b_{2}/a_{1}) & \cdots & p(b_{n}/a_{1})\\
p(b_{1}/a_{2}) & p(b_{2}/a_{2}) & \cdots & p(b_{n}/a_{2})\\
\vdots & \vdots & \ddots & \vdots\\
p(b_{1}/a_{n}) & p(b_{2}/a_{n}) & \cdots & p(b_{n}/a_{n})
\end{array}\right]
\end{eqnarray*}
The conditional probability transform matrix should be implemented by encryption matrix. When encryption matrix make
$P(\xi_A)$ and $P(\xi_B)$ independently of each other, our goal that $\mathbf{s}$ can not be obtained from $\mathbf{s}'$ could be achieved.
Although $P(\xi_B)$ can be any probability distribution, we should carefully choose a probability distribution to facilitate our implementation.
In this work, we set $P(\xi_B)$ as discrete equivalent probability distribution, namely, $p(b_i)=p(a_j/b_i)=1/n$ for each $i$ and $j$. Therefore, $\Gamma$ can be expressed as
\begin{eqnarray*}
Gamma  &=& \left[\begin{array}{cccc}
1/n & 1/n & \cdots & 1/n\\
1/n & 1/n & \cdots & 1/n\\
\vdots & \vdots & \ddots & \vdots\\
1/n & 1/n & \cdots & 1/n
\end{array}\right]
\end{eqnarray*}
According to the above $P(\xi_B)$ and $\Gamma$, $A$ should implement all the nonzero elements of $\mathbf{s}$ with uniform distribution for $\mathbf{s}'$.

\subsubsection{ Encryption Matrix Implementation}
In this part, we give our encryption matrix $A$ implementation based on above analysis.
We exploit random perturbation  identity matrix as $A$ to implement the function of conditional probability transformation matrix $\Gamma$. The definition of $A$ is
\begin{equation}\label{e_def_A}
    A=\pi(I)
\end{equation}
where $I$ is identity matrix. $\pi(\cdot)$ is random perturbation function with the same probability, which is equivalent to randomly perturbing the rows of $I$. According to this definition, we know that $A$ is an invertible matrix. Meanwhile, $A$ has encryption function because it could implement  all the elements of $\mathbf{s}$ and $\mathbf{s}'$  to be independent to each other. Since $B$ equals to $\Psi^{-1}A\Psi$ and $\mathbf{s}=B\mathbf{s}'$, $B$ only perturbs the element location of $\mathbf{s}$, and the sparsity of $\mathbf{s}$ is not changed. So our designed encryption matrix satisfies three conditions of section 3.2.
If all the elements of $\mathbf{s}$ are perturbed by $A$, we only need to implement the equal probability random perturbation for each element. In fact, $A$ could be easily implemented by a random seed, which is also considered as the key between the sampling device and the end user.

To quantify the privacy protection level more effectively, and make a more detailed description on security considerations,  we  next introduce the concept of $k$-secure.
\begin{definition}
The encryption matrix $A$ is $k$-secure if the number of permutation rows is $k$. It is denoted by
\begin{equation}\label{e_sd}
    A_k=\pi_k(I)
\end{equation}
$A_k$ is called $k$-secure.
\end{definition}

If $k=0$, $A_0$ is equivalent to $I$, i.e., no encryption operation is carried out. On the other hand, we achieve the highest security level by setting $k=n$ , i.e., perturb all elements in original sparse signal. 
Consider that $\mathbf{s}$ is sparse, in order to implement non-zero elements of $\mathbf{s}'$ with uniform distribution, we should select more non-zero elements to carry out perturbation operations. If the non-zero elements distribution of $\mathbf{s}$ is known, the same probability can be used as perturbation probability for each element. Otherwise, uniform random selection is adopted for each element. 

Since $A$ is random perturbation matrix of $I$, $\Phi A$ is equivalent to perturb the column of $\Phi$. The image compression and encryption could be implemented simultaneously. 

\subsection{System Design}
\textbf{Encryption and compression component:}
Sampling device is responsible for sampling, compression and encryption operations. We mainly consider compression and encryption after image signal sampling. In encryption and compression component, we exploit the common measurement matrix, Gaussian random matrix $\Phi$, as our measurement matrix. The encryption and compression process is calculate as $\mathbf{y}'=\Phi A \mathbf{f}$. Due to $A$ is random perturbation identity matrix by identity matrix $I$, it can be implemented by a random seed. $\Phi A$ is equivalent to perturb the column of $\Phi$.

\textbf{Cloud component:}
In eCIS, cloud component is responsible for storing user's compressed image signal $\mathbf{y}'$ and decoding the encryption sparse signal $\mathbf{s}'$. If there is no user's request, cloud only stores the user's image signal $\mathbf{y}'$  and $\Phi$. If the user requests an image signal, it sends $\Psi$ to the cloud. The cloud receives the user's request and $\Psi$, and carries out CS decoding operation according to Eq. \ref{e_cry_cs_dec}. After decoding the encryption sparse signal $\mathbf{s}'$, the cloud sends it to the end user. Considering that cloud could be publicly accessible, the attacker can obtain $\mathbf{s}'$ and $\Psi$. Although the attacker can obtain the user's data, it can only recover the encryption image signal via $\Psi \mathbf{s}'$ ($A\mathbf{f}=\Psi \mathbf{s}'$).
According to our encryption matrix, cloud can implement both storage and decoding computation functions allowing significant reduction of resource consumption on the sampling device and end user.

\textbf{End user component:}
The end user is responsible for decrypting and recovering the original image.  If the end user receives $\mathbf{s}'$, it carries out decryption and recovery operations. According to Eq. \ref{e_cry_cs_enc3}, we know $A\mathbf{f}=\Psi \mathbf{s}'$. Therefore, the recovered signal is given by
\begin{equation}\label{e_eusr_rec}
    \mathbf{f}=A^{-1}\Psi \mathbf{s}'
\end{equation}
Since $A$ is random perturbation identity matrix of identity matrix $I$, $A^{-1}$ is equivalent to the transpose form of $A$. Therefore, $\mathbf{f}=A^T\Psi \mathbf{s}'$. $A^T\Psi$ is equivalent to perturbing the rows of $\Psi$. The image recovery and decryption could be easily implemented, and eCIS does not need  complicated decryption operation compared with transformed-based image recovery.

\section{Theoretical Analysis}
\subsection{Security Issue}\label{s_sa}
According to our encryption matrix definition as shown in Eq.\ref{e_sd}, if the attacker wants to obtain the original sparse signal and $A$ is $k$-secure, he should investigate all possible  arrangements as
\begin{equation*}
    C(n,k)\times n(n-1)\cdots (n-k+1)= C(n,k)\cdot \frac{n!}{(n-k)!}
\end{equation*}
where $C(n,k)=\frac{n!}{(n-k)!k!}$. If the attacker also does not know the value of $k$, the number of possible combinations is
\begin{equation*}
    \sum_{k=1}^{n}(C(n,k)\cdot \frac{n!}{(n-k)!})
\end{equation*}
Suppose that the attacker knows the value of $k$, the probability to successfully recover  $\mathbf{s}$, $P_{suc}$, can be calculated according to Stirling's approximation \cite{feller1968stirling}
\begin{eqnarray}\label{e_cnk}
    P_{suc} & = & \frac{1}{C(n,k)\times \frac{n!}{(n-k)!}}= \frac{1}{C(n,k)\times C(n,k)\times k!}\nonumber\\
    & \leq & \frac{1}{e(en^2/k)^k} \leq  \frac{1}{e(en)^k} =  e^{-(k\log n+k+1)}
\end{eqnarray}
If the recovery probability is required to be less than $\beta$,  $k$ should satisfy the following inequality:
\begin{equation*}
    k\geq \left\lceil\frac{-\log\beta-1}{\log n +1}\right\rceil
\end{equation*}

The signal $\mathbf{s}$, however,  is sparse, the attacker may not need to try too many perturbations of zero elements.
To simplify the problem analysis, we do not consider the nonzero element probability distribution of $\mathbf{s}$. We assume that there are $l$ perturbed elements among $\mathbf{s}$ and $\mathbf{s}'$ to be considered. According to Eq.\ref{e_cnk}, the probability of successfully recovered $\mathbf{s}$, $P_{suc}$, is given by
{\footnotesize
\begin{eqnarray}\label{e_cnt}
    P_{suc} & = & \frac{1}{C(n,l)\times \frac{n!}{(n-l)!}}
     \leq  \frac{1}{e(en)^l}
    =  e^{-(l\log n+l+1)}
\end{eqnarray}}

\begin{figure}
  \begin{center}
  \includegraphics[scale=0.35]{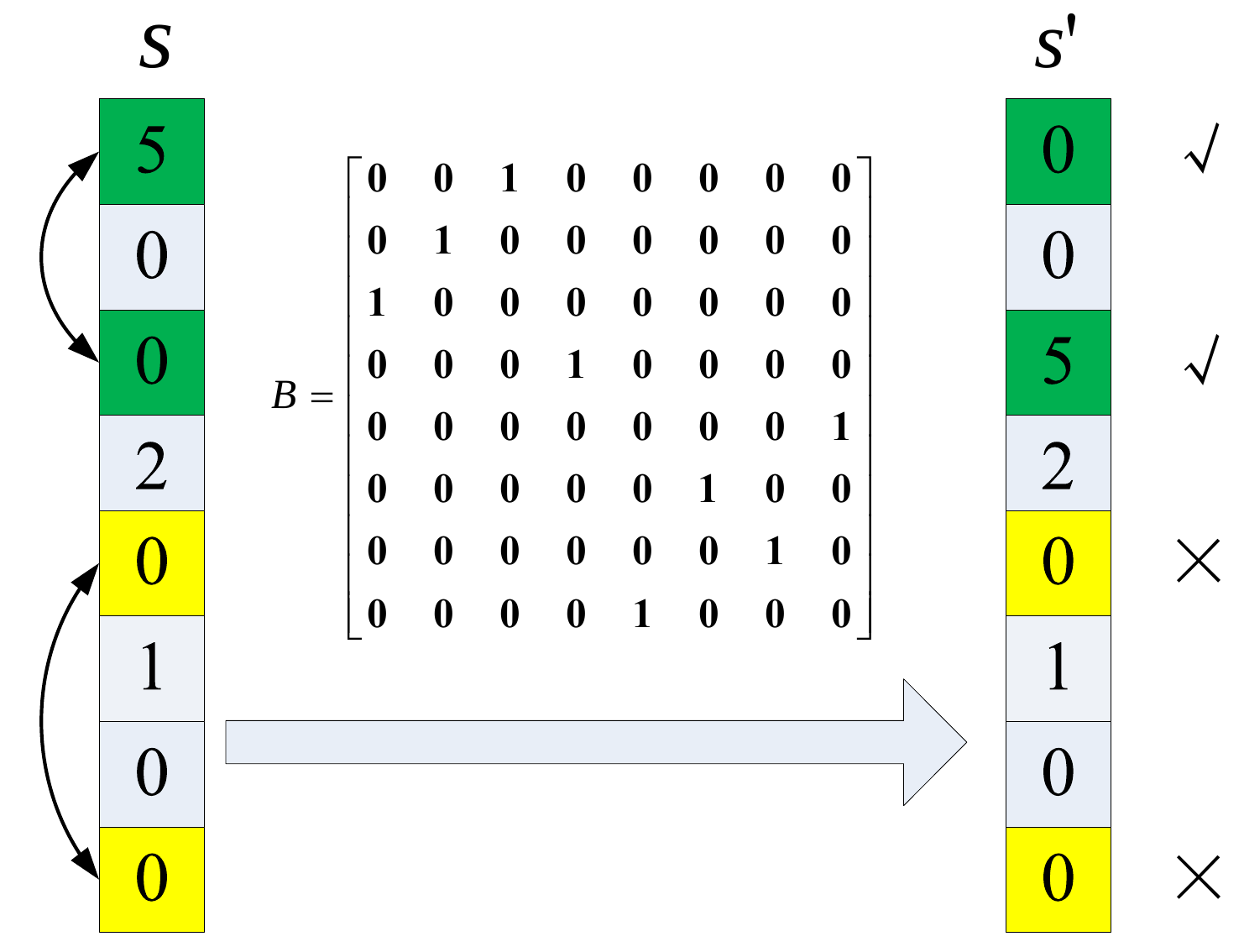}\\
  \caption{{\footnotesize Example of $s$ and $s'$ under $A$.}}\label{f_examp_pb}
  \end{center}
\end{figure}

Nextly, we calculate the value of $l$. We suppose that $\mathbf{s}$ is $t$-sparse and $A$ is $k$-secure. 
First of all, we  look at what circumstance the attacker does not need to consider.  For example, Fig.\ref{f_examp_pb} shows that $4$ elements are perturbed between $\mathbf{s}$ and $\mathbf{s}'$, namely, $A$ is $4$-secure. The attacker only needs to guess the first two elements $0$ and $5$, and does not consider the other two perturbed $0$ elements. In other words, the attacker does not consider the case of two zero elements perturbation. Since the number of selection zero elements is decided by the non-zero element distribution, we conduct the security analyses with uniform and nonuniform distribution of non-zero elements in $\mathbf{s}$.

\subsubsection{Uniform Distribution}
When the nonzero element follows uniform distribution in $\mathbf{s}$, each zero element of $\mathbf{s}$ has the same perturbation probability. The probability of two zero elements perturbation is
\begin{equation*}
\frac{n-t}{n}\cdot\frac{n-1-t}{n-1}=\frac{(n-t)(n-t-1)}{n(n-1)}
\end{equation*}
Therefore, the average number of zero elements which is not to be considered is
\begin{equation*}
k\cdot\frac{(n-t)(n-t-1)}{n(n-1)}
\end{equation*}
The number of perturbed elements which needs to be considered, $l$,  is
\begin{equation*}
    l=k\cdot(1-\frac{(n-t)(n-t-1)}{n(n-1)})=\frac{kt(2n-t-1)}{n(n-1)}
\end{equation*}

Given that $t(2n-t-1)/(n(n-1))$ is a constant, the number of perturbed elements $l$ can be denoted by $\alpha\cdot k$, where $\alpha=t(2n-t-1)/(n(n-1))$ and $0<\alpha<1$.
According to Eq.\ref{e_cnt},  $P_{suc}$ can be calculated as
\begin{eqnarray} \label{e_cnl}
    P_{suc} & \leq & \frac{1}{e(en)^{\alpha k}} = e^{-(\alpha k(\log n+1)+1)}
\end{eqnarray}

Eq.\ref{e_cnl} indicates  that the complexity to recover  the original signal is $O(n^{\alpha k\log n})$. 
If the recovery probability is required to be less than $\beta$, $k$ should satisfy the following inequality:
\begin{equation*}
    k\geq \left\lceil\frac{-\log\beta-1}{\alpha (\log n+1)}\right\rceil=\left\lceil\frac{n(n-1)(-\log\beta-1)}{t(2n-t-1) (\log n+1)} \right\rceil
\end{equation*}

\begin{figure*}[t]
 \begin{center}
  \includegraphics[scale=0.25]{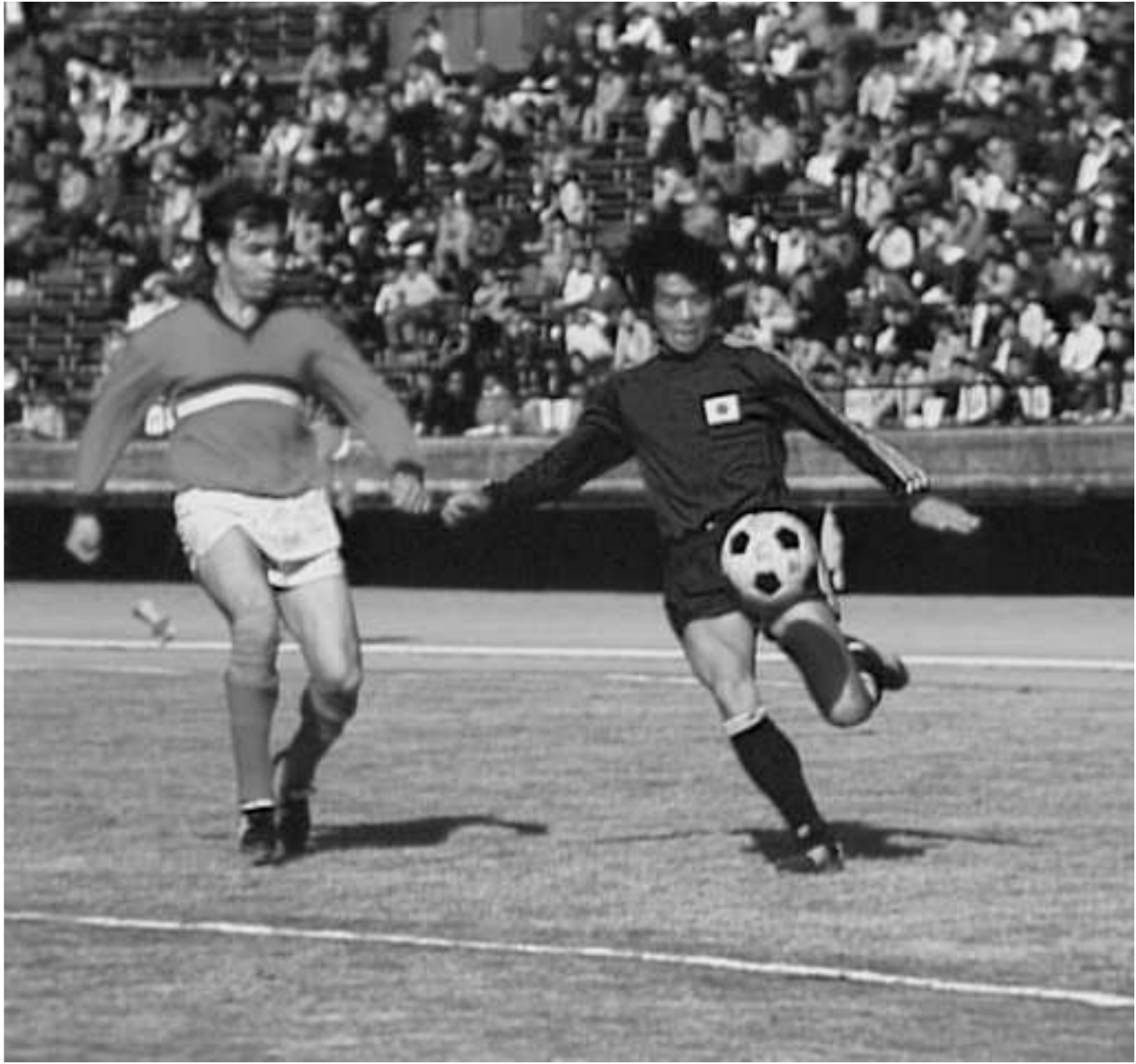}
  \includegraphics[scale=0.25]{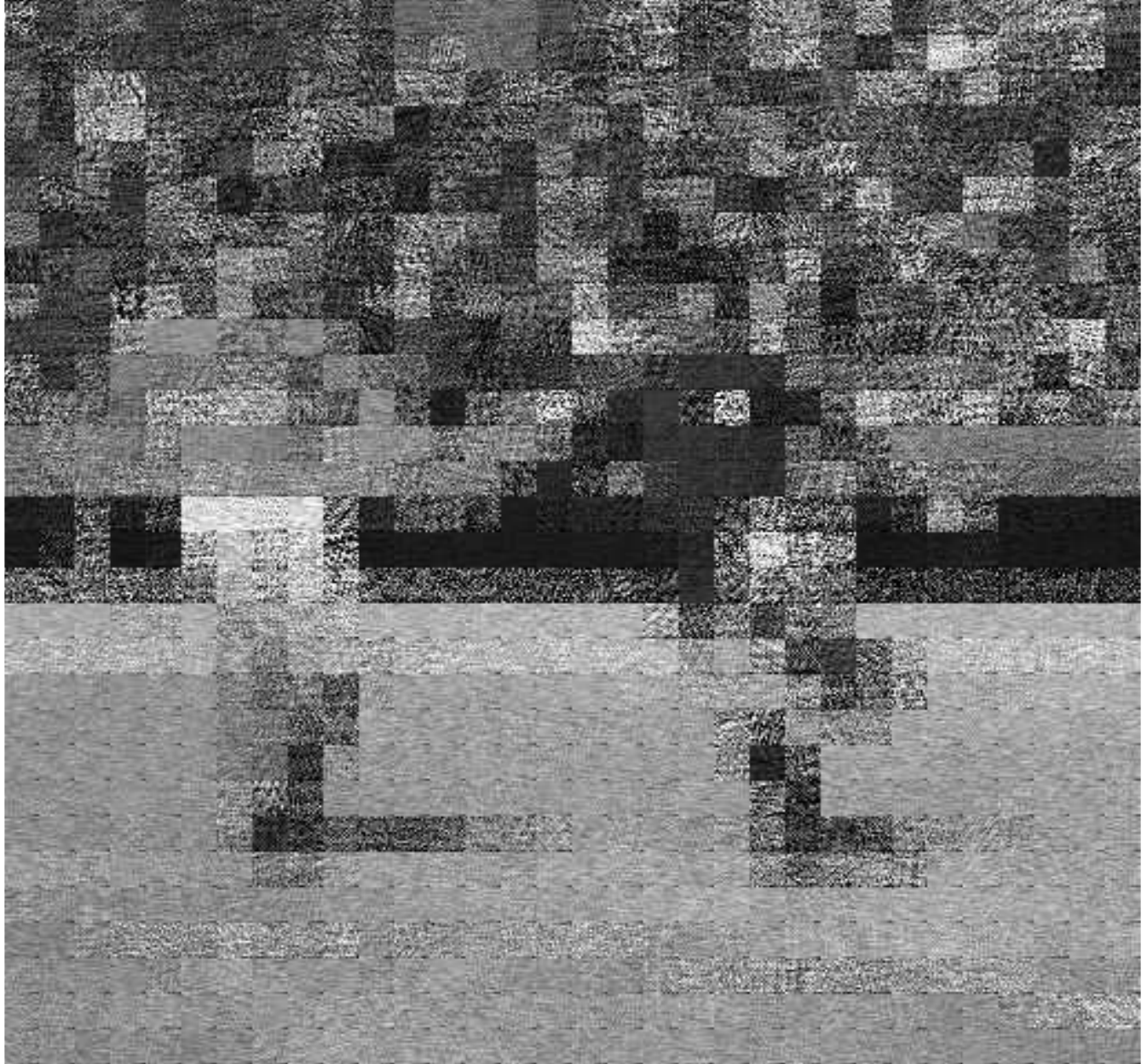}
  \includegraphics[scale=0.25]{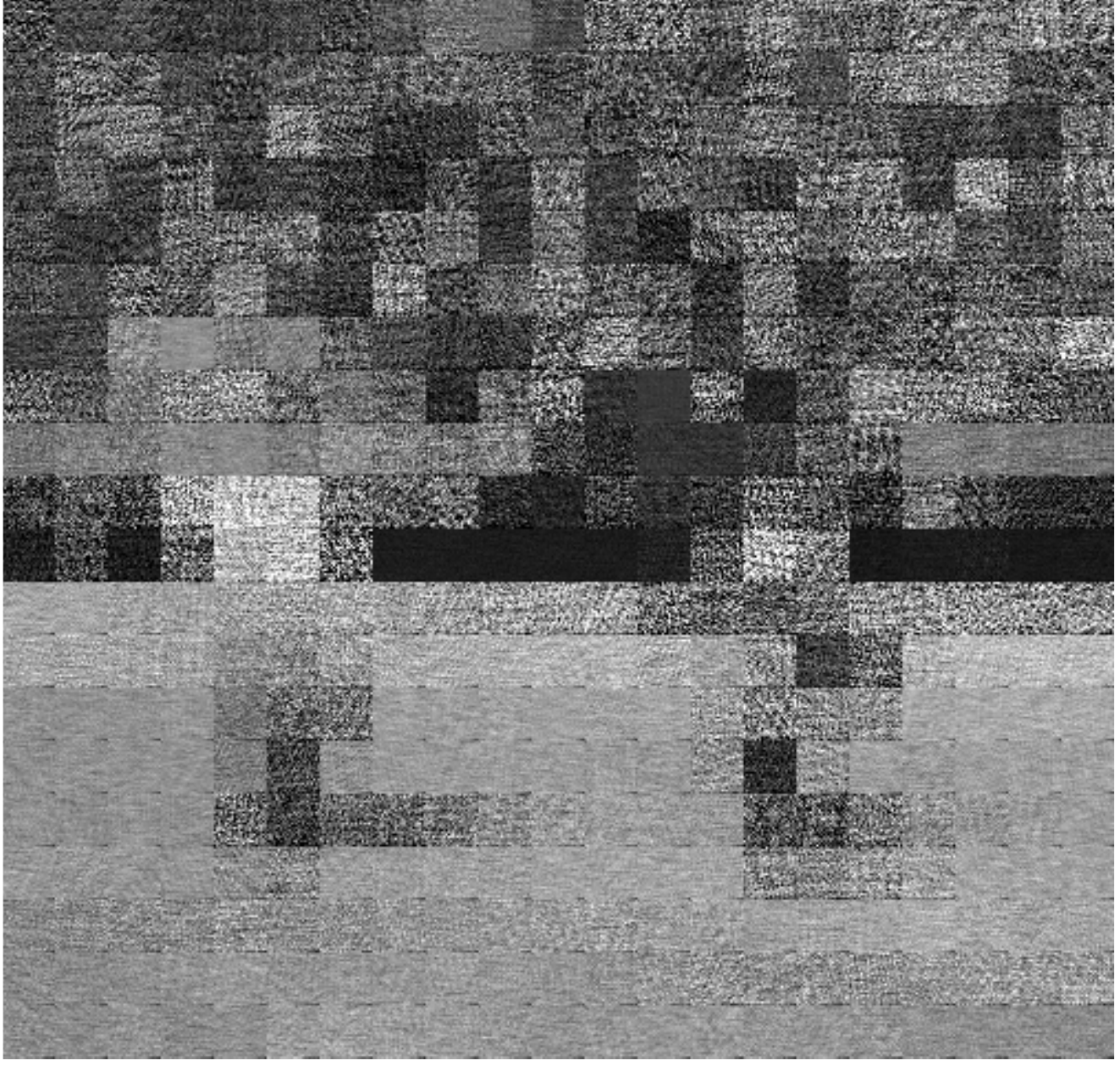}
  \includegraphics[scale=0.25]{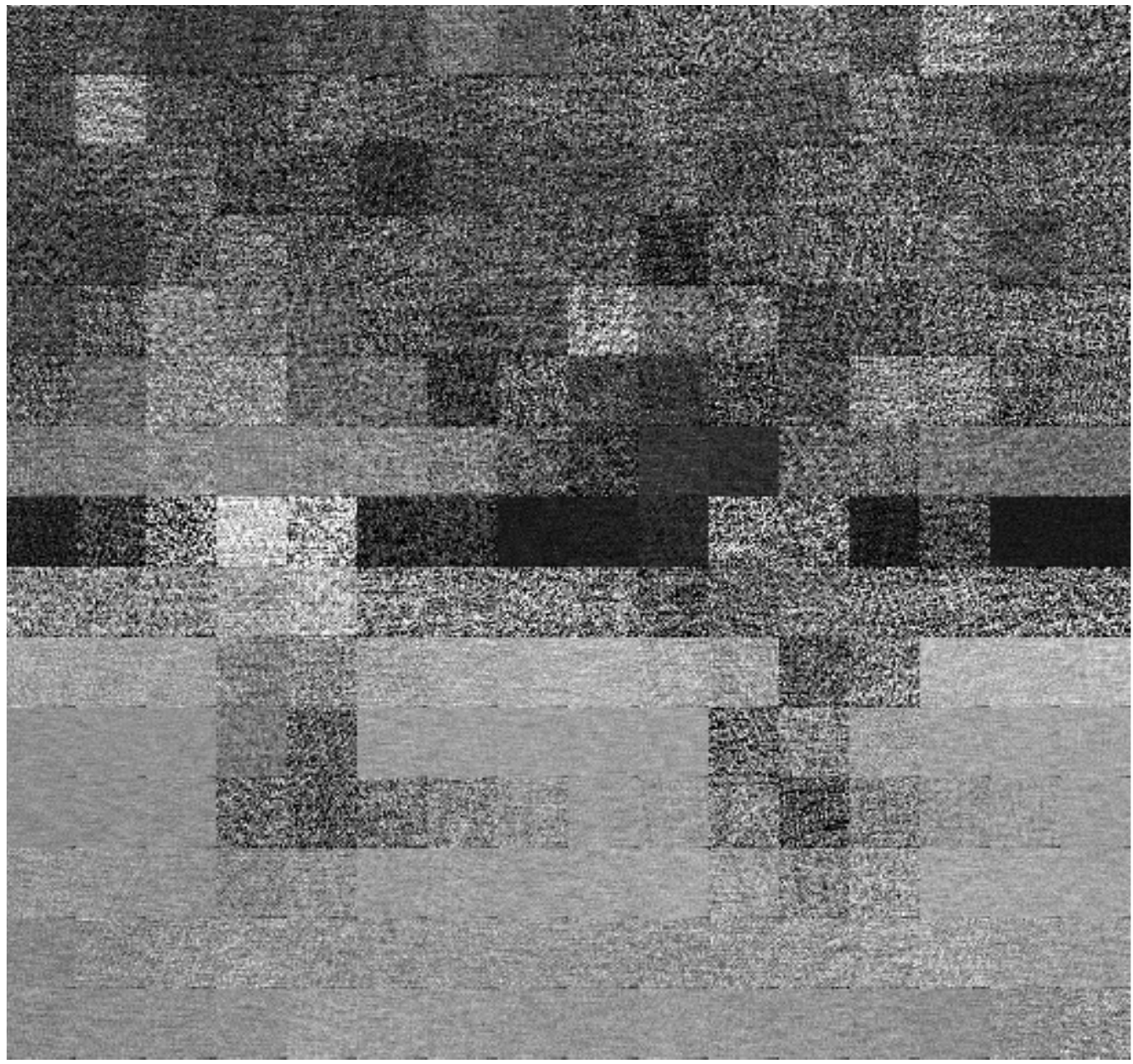}
  \includegraphics[scale=0.25]{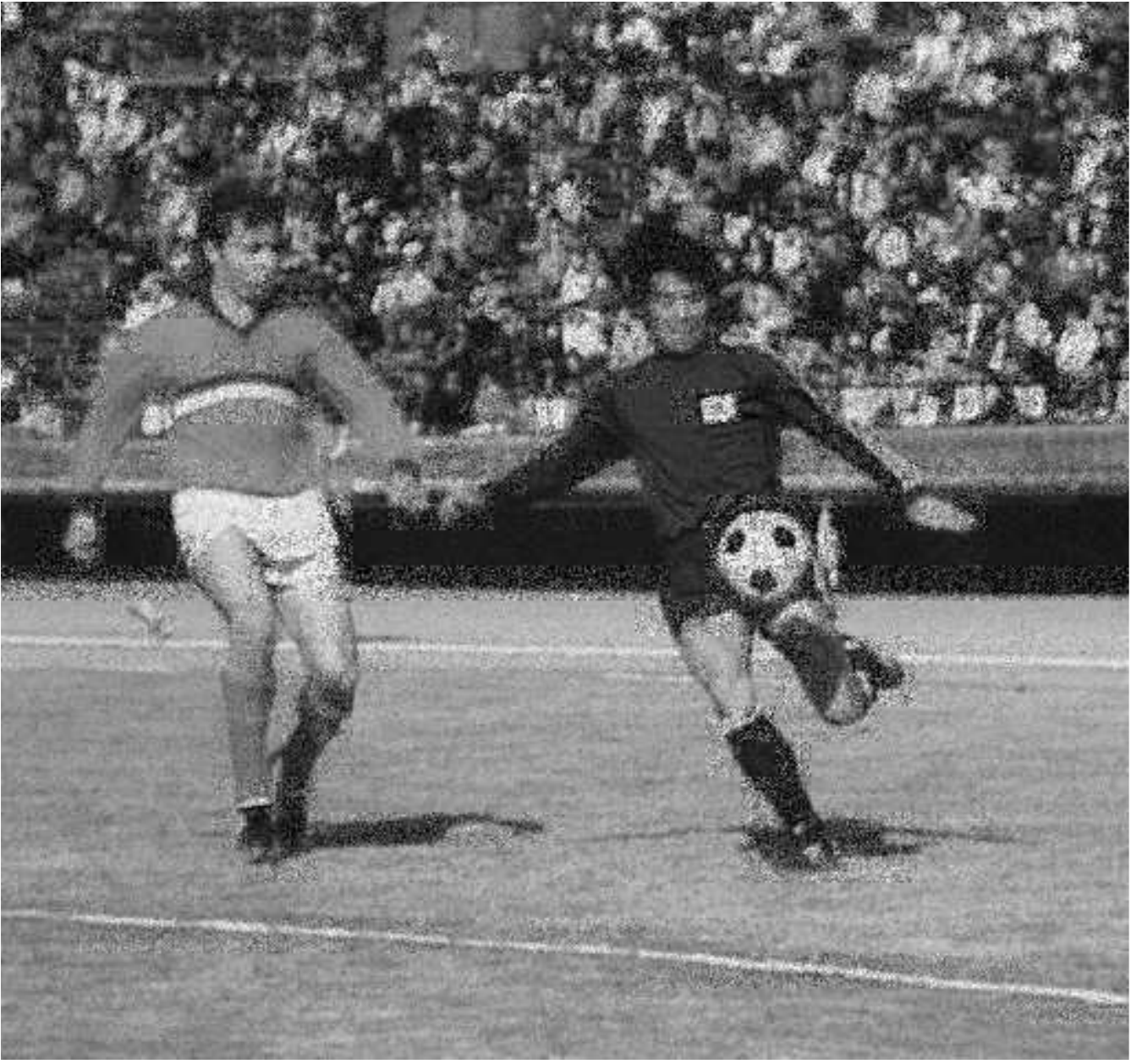}  \\
 {\footnotesize (a1) ~~~~~~~~~~~~~~~~~~~~~~~~~~~~~(b1)~~~~~~~~~~~~~~~~~~~~~~~~~~~~~ (c1)~~~~~~~~~~~~~~~~~~~~~~~~~~~~~(d1)~~~~~~~~~~~~~~~~~~~~~~~~~~~~~(e1) }\\
  \includegraphics[scale=0.27]{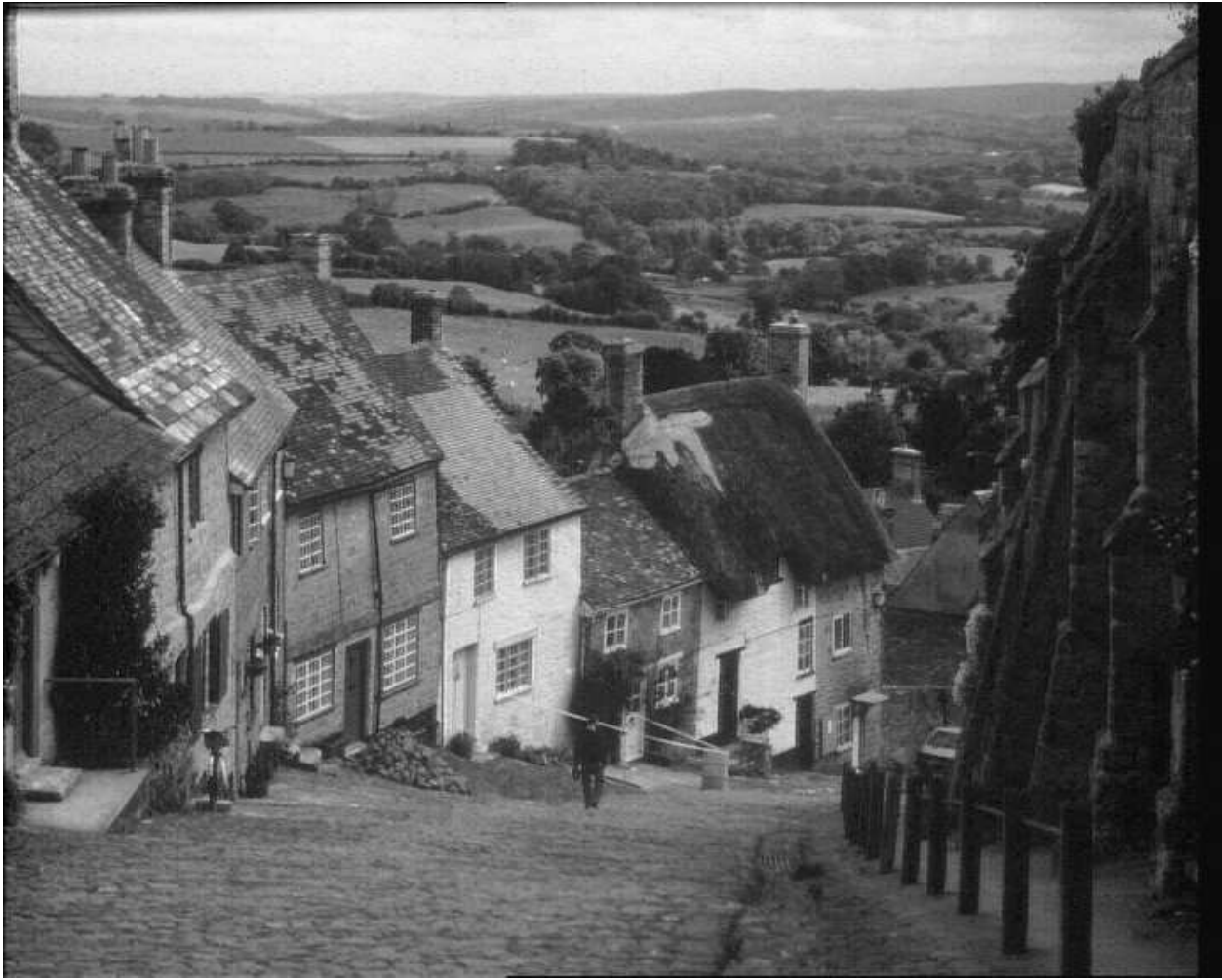}
  \includegraphics[scale=0.27]{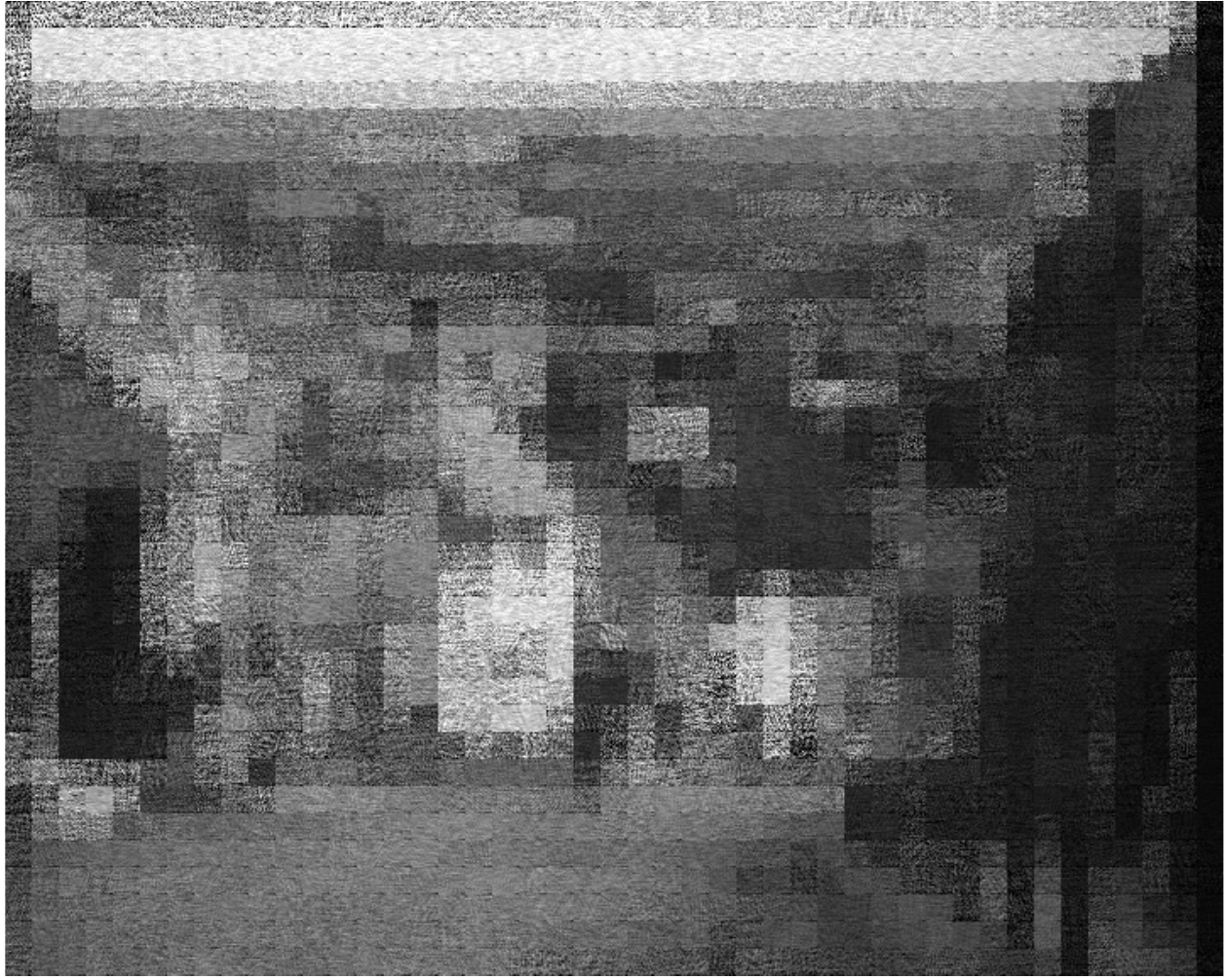}
  \includegraphics[scale=0.27]{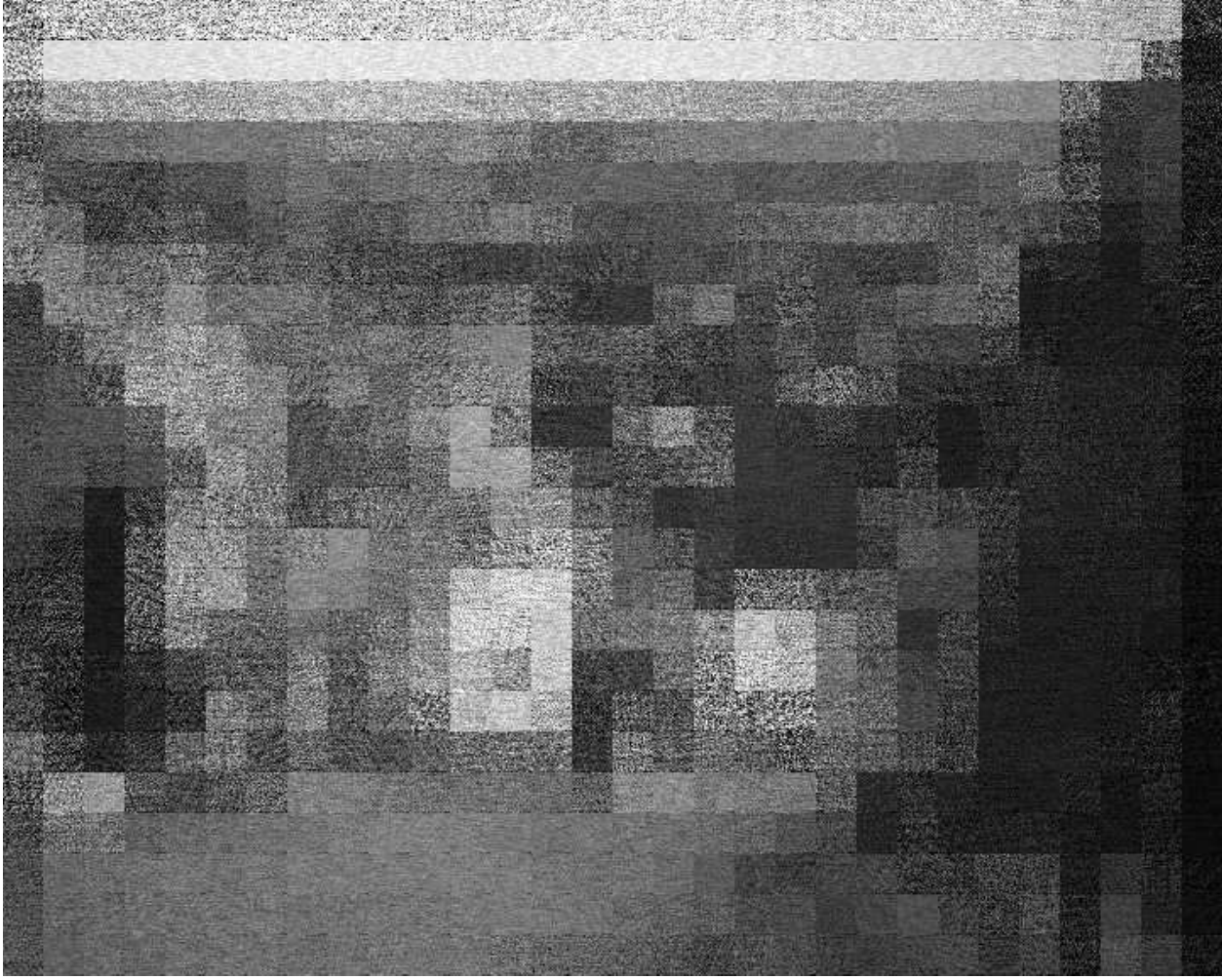}
  \includegraphics[scale=0.27]{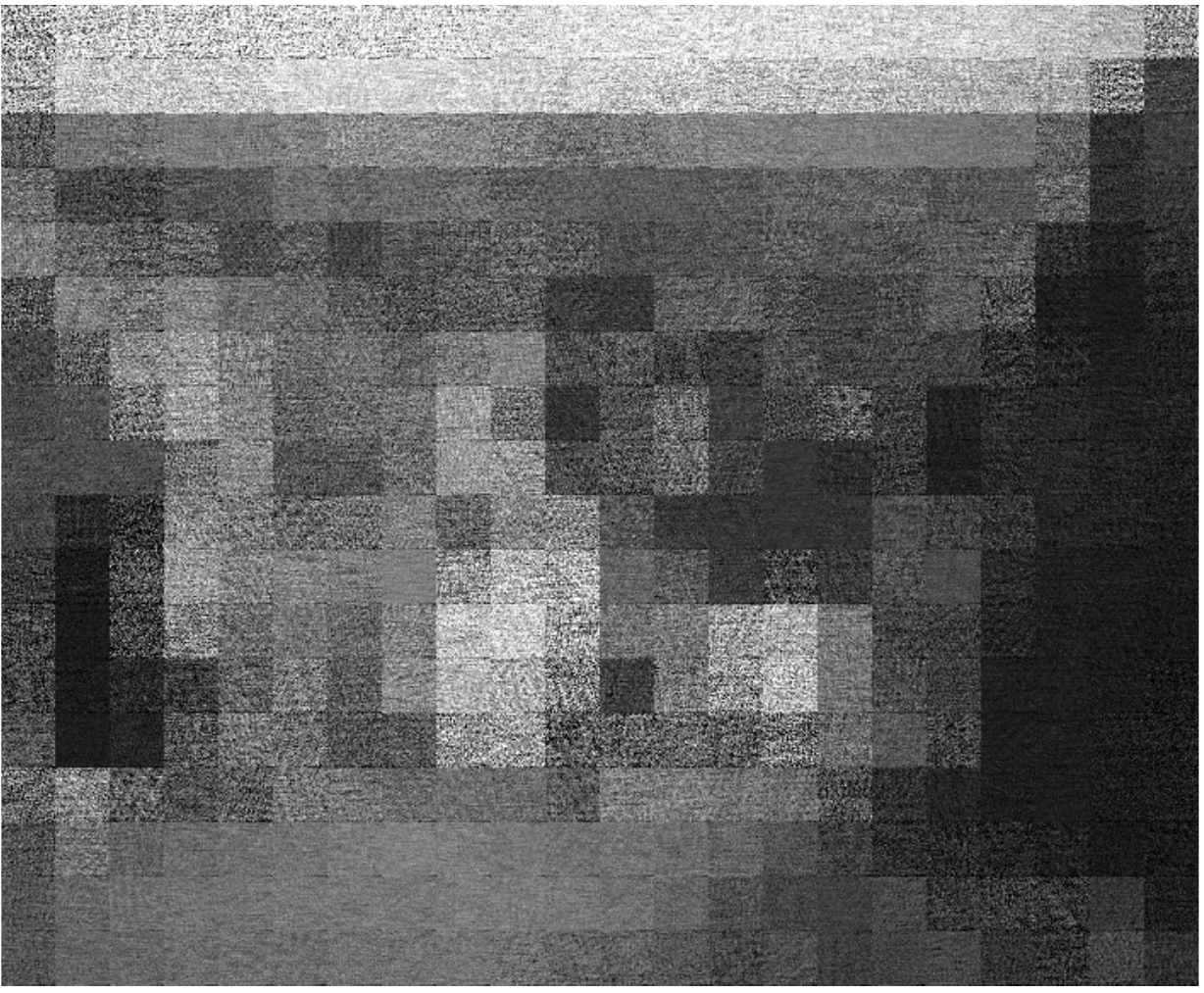}
  \includegraphics[scale=0.27]{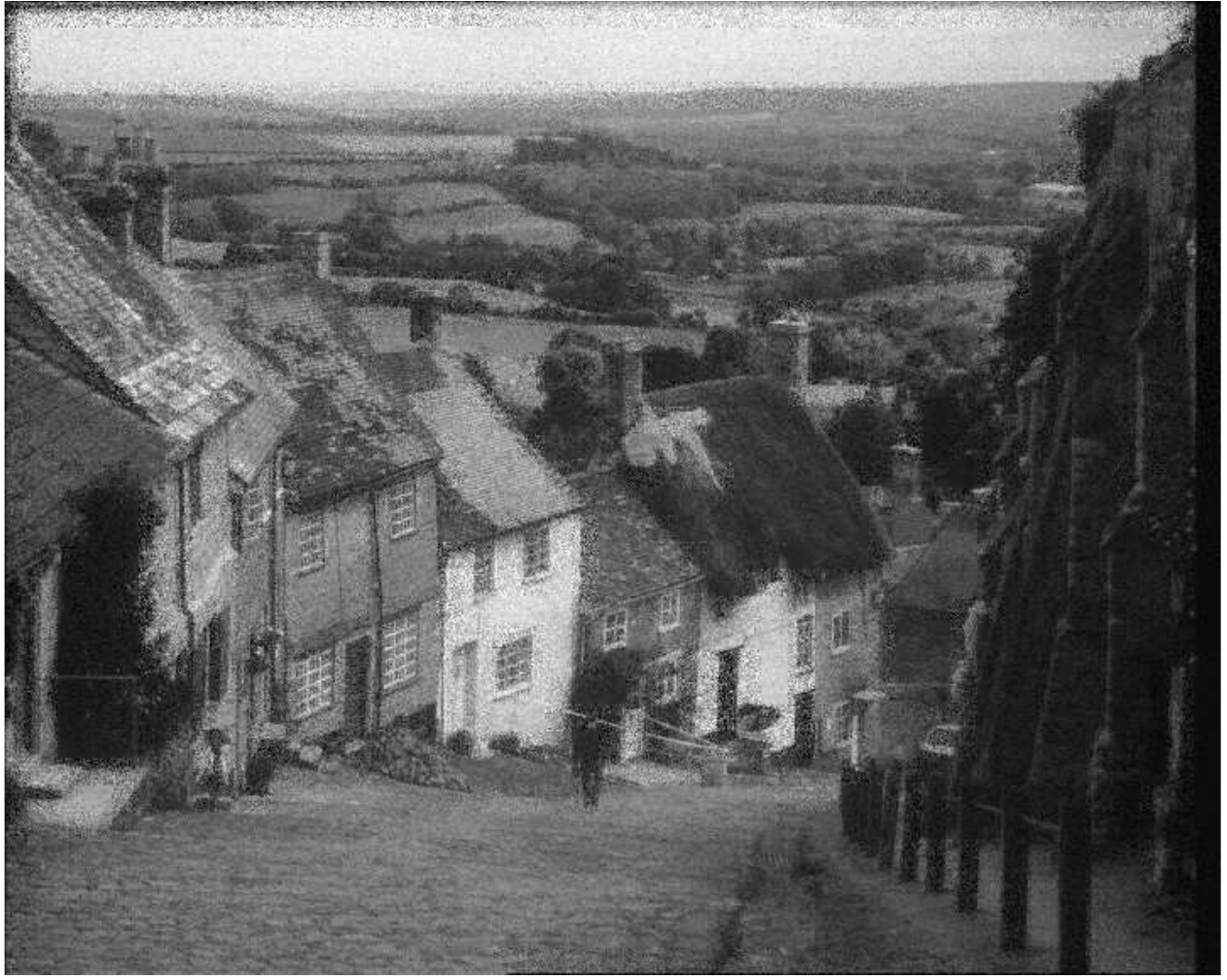}  \\
   {\footnotesize (a2) ~~~~~~~~~~~~~~~~~~~~~~~~~~~~~(b2)~~~~~~~~~~~~~~~~~~~~~~~~~~~~~ (c2)~~~~~~~~~~~~~~~~~~~~~~~~~~~~~(d2)~~~~~~~~~~~~~~~~~~~~~~~~~~~~~(e2) }\\
 \end{center}
  \caption{{\footnotesize Cloud-assisted image encryption with different block sizes. (a1) and (a2). Original images. (b1) and (b2). $16\times 16$ pixels of block size. (c1) and (c2). $24\times 24$ pixels of block size. (d1) and (d2).   $32\times 32$ pixels of block size. (e1) and (e2). Recovery image of end user.}}\label{f_se_diff_block_size}
\end{figure*}

\subsubsection{Nonuniform Distribution}
We next study the case when  the nonzero element follows nonuniform distribution in $\mathbf{s}$, we denote the distribution probability is $P(\mathbf{s})$ and the probability of the $i^{th}$ location is $p(s_i)$. We also assume that the element selection probability is $q(s_i)$. The expected probability  for choosing an non-zero element in one selection is
 $\sum_{i=1}^n p(s_i)\cdot q(s_i)$. The probability that we choose  two zero elements for a perturbation is  $(1-\sum_{i=1}^n $ $p(s_i)q(s_i))^2$. 
To reduce the number of selected zero elements, $\sum_{i=1}^n p(s_i)q(s_i)$ should be as large as possible. According to principle of the maximum, $\sum_{i=1}^n p(s_i)q(s_i)$ achieves a maximum value  if $q(s_i)$ is equal to $p(s_i)$. In other words, the expected number of selected zero element is minimized if the selection strategy is the same as the nonzero elements distribution.

Since $q(s_i)$ is equivalent to $p(s_i)$,  the number of perturbed elements which the attacker needs to consider  can be expressed as
\begin{equation*}
   l= k\cdot(1-(1-\sum_{i=1}^n p(s_i)^2)^2)
\end{equation*}
Then,  $P_{suc}$ is given by
\begin{eqnarray}
    P_{suc} & \leq &  e^{-((1-(1-\sum_{i=1}^n p(s_i)^2)^2) k(\log n+1)+1)}
\end{eqnarray}
If the recovery probability is required to be less than $\beta$, the value $k$ should satisfy the following inequality
\begin{equation*}
    k\geq \left\lceil\frac{(-\log\beta-1)}{((1-(1-\sum_{i=1}^n p(s_i)^2)^2)\log n+1)} \right\rceil
\end{equation*}

When the distribution of nonzero elements is unknown, we consider the perturbed elements as uniform random strategy, and $q(s_i)$ is equivalent to  $1/n$. Therefore,
\begin{eqnarray}\label{e_cn2}
    P_{suc} & \leq &  e^{-((1-(1-\frac{1}{n}\sum_{i=1}^n p(s_i))^2) k(\log n+1)+1)} \nonumber \\
    & = & e^{-(\frac{2n-1}{n^2} k(\log n+1)+1)}
\end{eqnarray}
Eq.\ref{e_cn2} indicates  that the complexity to recover  the original signal is $O(n^{(k\log n)/n})$.
If the recovery probability is required to be less than $\beta$, the value of $k$ should satisfy the following inequality.
\begin{equation*}
    k\geq \left\lceil\frac{(-n^2\log\beta-1)}{((2n-1)\log n+1)} \right\rceil
\end{equation*}

\textbf{Remark:} According to the above security analysis, the user can adjust the security level $k$ according to his privacy requirement. For $k$-secure requirements, the computational complexity for attacker is $O(n^{(tk\log n)/n})$ when uniform distribution is concerned for non-zero elements, and $O(n^{(k\log n)/n})$ when non-uniform distribution is concerned.

\subsection{Overhead Analysis}
\textbf{Computational Complexity:}
In this part, we analyze computational complexity for each component of eCIS. For compression and encryption operations, it calculates $\mathbf{y}'=\Phi A\mathbf{f}$. Since $A$ is random perturbation matrix of identity matrix $I$, calculating $\Phi A$ is equivalent to perturbing the column of $\Phi$ whose  complexity is $O(mn)$. If the sampling device only carries out CS-based compression without encryption, the  complexity is also $O(mn)$. If the sampling device takes LP transform outsourcing \cite{wang2014privacy}, its complexity is $O(n^\theta+mn)$ ($2<\theta<3$).

For end user component, the decryption and recovery processes are $\mathbf{f}=A^{-1}\Psi \mathbf{s}'$, and $A^{-1}$ is equivalent to $A^{-T}$. Computing $A^T\Psi$ only requires to perturb the rows of $\Psi$ , thus the complexity of end user component is  $O(n^2)$. Assume  the end user computes  the CS solution via Eq. \ref{e_cs_dec} without outsourcing it to the cloud, the computation cost would be $O(n^3)$ (e.g \cite{OMP}).

At the cloud side, the computation  cost is $O(n^3)$ which is equivalent to the complexity of  solving  Eq. \ref{e_cry_cs_dec}. 
In  eCIS, we shift the $\ell_1$ optimization from the end user to the cloud without adding extra computation cost. More importantly, eCIS provides privacy protection  compared with the original cloud-assisted CS decoding.

\textbf{Communication Cost:} In eCIS, the transmission cost between the sampling device and cloud is measured by  the number of CS measurements that are transmitted. Since our scheme does not change the sparsity of the original image signal,  the transmission cost remains the same as that of the nonencryption CS-based image data compression. Although the scheme proposed in  \cite{wang2014privacy} has the same transmission cost between the sampling device and cloud, the transmission cost between cloud and end user is very high because the cloud needs to send the encrypted original image data to the end user. In our scheme, cloud only sends the encrypted sparse signal to the end user allowing great reduction of  the transmission cost. For example, when  the compression ratio is no less than  $50\%$, our scheme can reduce half of the transmission cost compared with the scheme proposed in \cite{wang2014privacy}.


\section{Experiment}
In this section, we carry out extensive experiments to evaluate the performance of eCIS. In our experiment, we use different test sequences as the signal source of resource-constrained device. Our experiment is implemented via MATLAB, on a laptop with an Intel Core i5 CPU running at 1.6GHz and 4G RAM. Gaussian random matrix  is considered as our measurement matrix. Since the original image signal is not sparse in spatial domain, we select DCT basis as sparse representation basis because it is the most common sparse representation basis for image compression. In order to efficiently implement eCIS, the experimental images are divided into multiple the same size blocks. We also transform 2-dimension block pixel values into 1-dimension signal via row sequence to meet CS requirement.
Meanwhile, we evaluate the performances of our scheme form two aspects: effectiveness and computation overhead.

\begin{figure}[t]
 \begin{center}
  \includegraphics[scale=0.4]{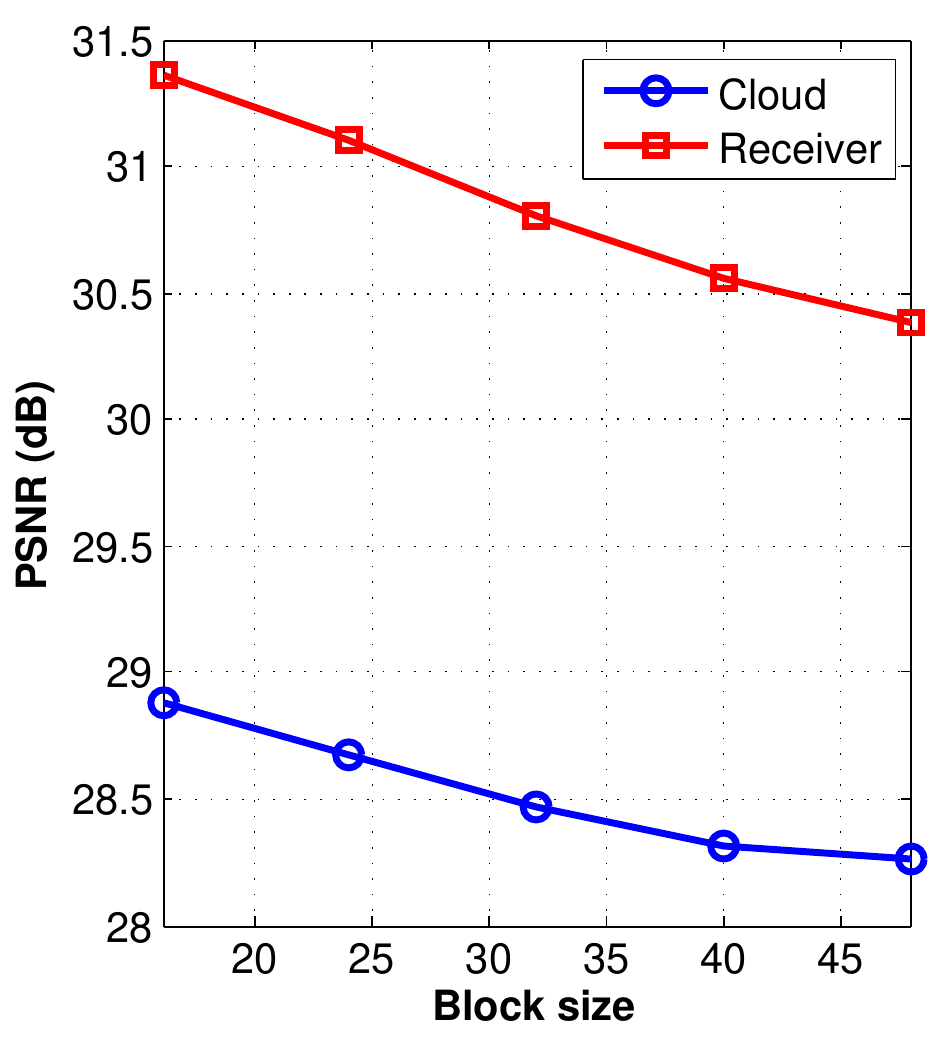}~~~~
  \includegraphics[scale=0.4]{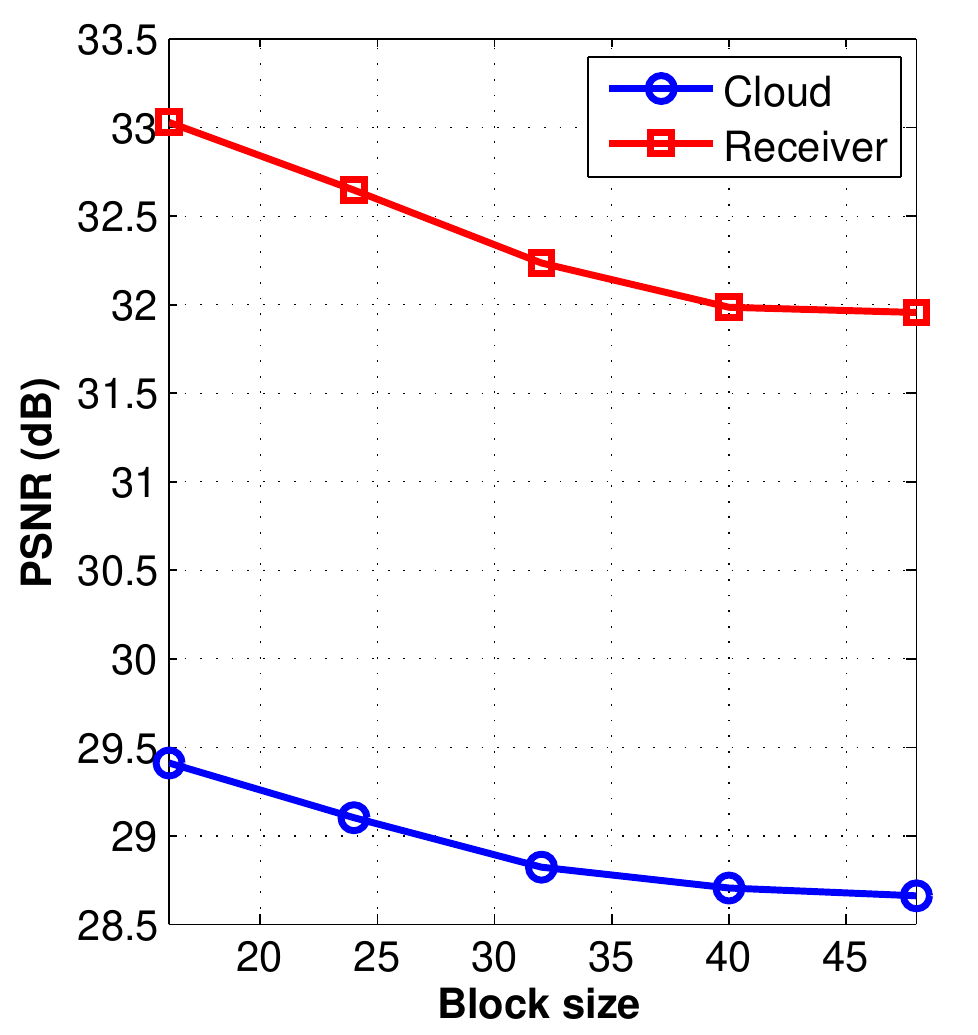}\\
  {\footnotesize ~~~~(a) ``Soccer".~~~~~~~~~~~~~~~~~~~~~~~~~(b) ``Goldhill".} \\
  \caption{{\footnotesize PSNR comparison of recovery image between cloud and end user with different block size. }}\label{f_psnr_diff_block_size}
 \end{center}
\end{figure}

\begin{figure}[t]
 \begin{center}
  \includegraphics[scale=0.23]{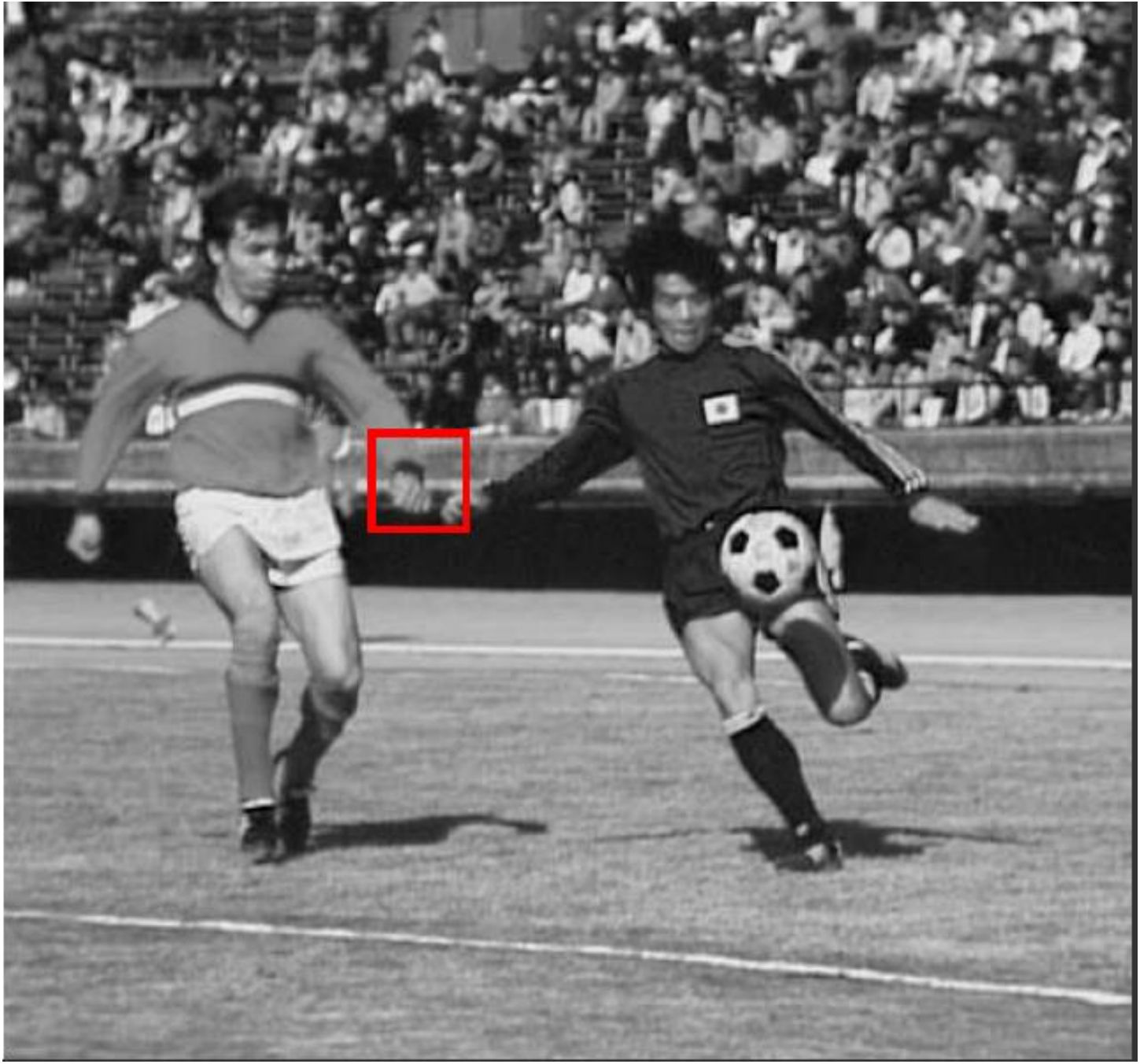} ~~~~~~~~
  \includegraphics[scale=2.9]{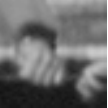}\\
  {\footnotesize (a)~~~~~~~~~~~~~~~~~~~~~~~~~~~~~~~~(b)}\\
  \includegraphics[scale=3.1]{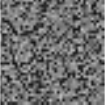} ~~~~~~~~
  \includegraphics[scale=0.6]{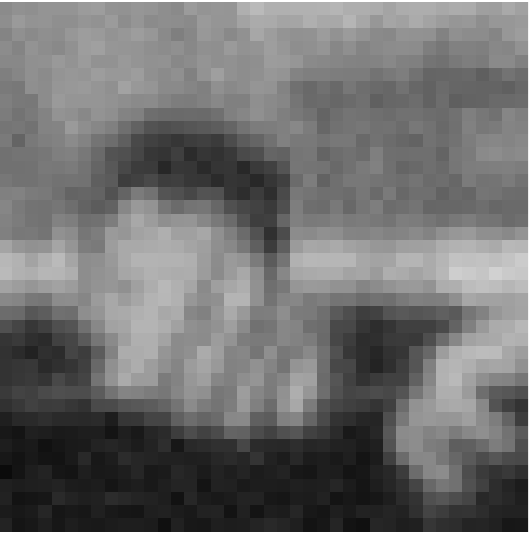}\\
  {\footnotesize (c)~~~~~~~~~~~~~~~~~~~~~~~~~~~~~~~~(d)} \\
  \caption{{\footnotesize Cloud-assisted image encryption of one block. (a) Original image and selected block. (b) $40\times 40$ pixels block. (c) Cloud recovery image with $n$-secure. (d) Recovery image of end user.}}\label{f_se_diff_1_block_size}
 \end{center}
\end{figure}

\subsection{Effectiveness Evaluation}
Our goal of effectiveness evaluation is to display image recovery performance of cloud and end user. We evaluate its performance from three aspects: different block size, different  security level, and adaptive region of interest (ROI) of test images. In our experiment, subjective visual effect and peak signal-to-noise ratio (PSNR) are considered as metrics to evaluate the quality of the recovery image. PSNR is defined as
\begin{equation*}
    PSNR=10\times \log_{10}\frac{255^2}{MSE}
\end{equation*}
where MSE is mean square error of gray scale pixel values between the original and recovery images.

\begin{figure*}[t]
 \begin{center}
  \includegraphics[scale=0.25]{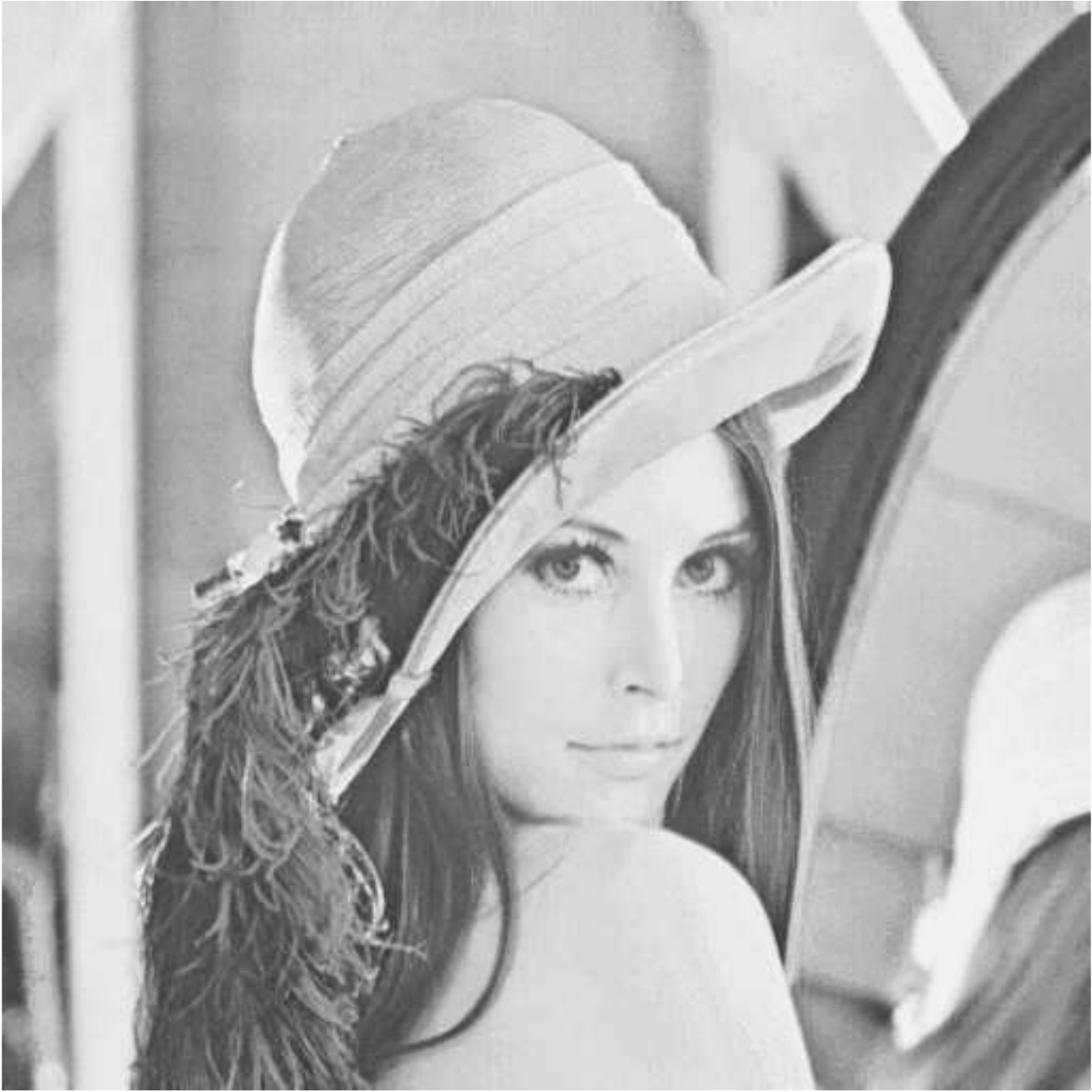}
  \includegraphics[scale=0.25]{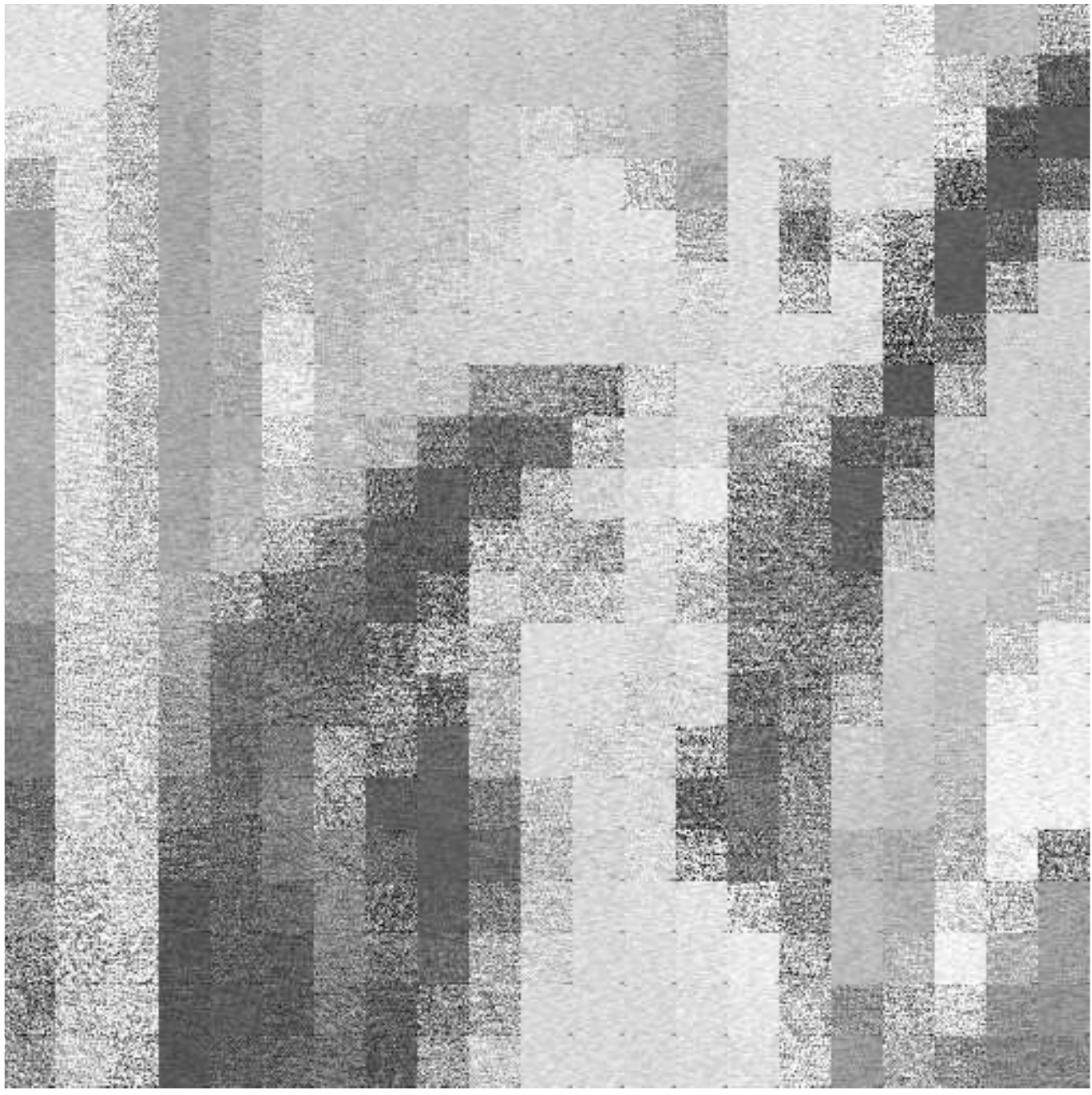}
  \includegraphics[scale=0.25]{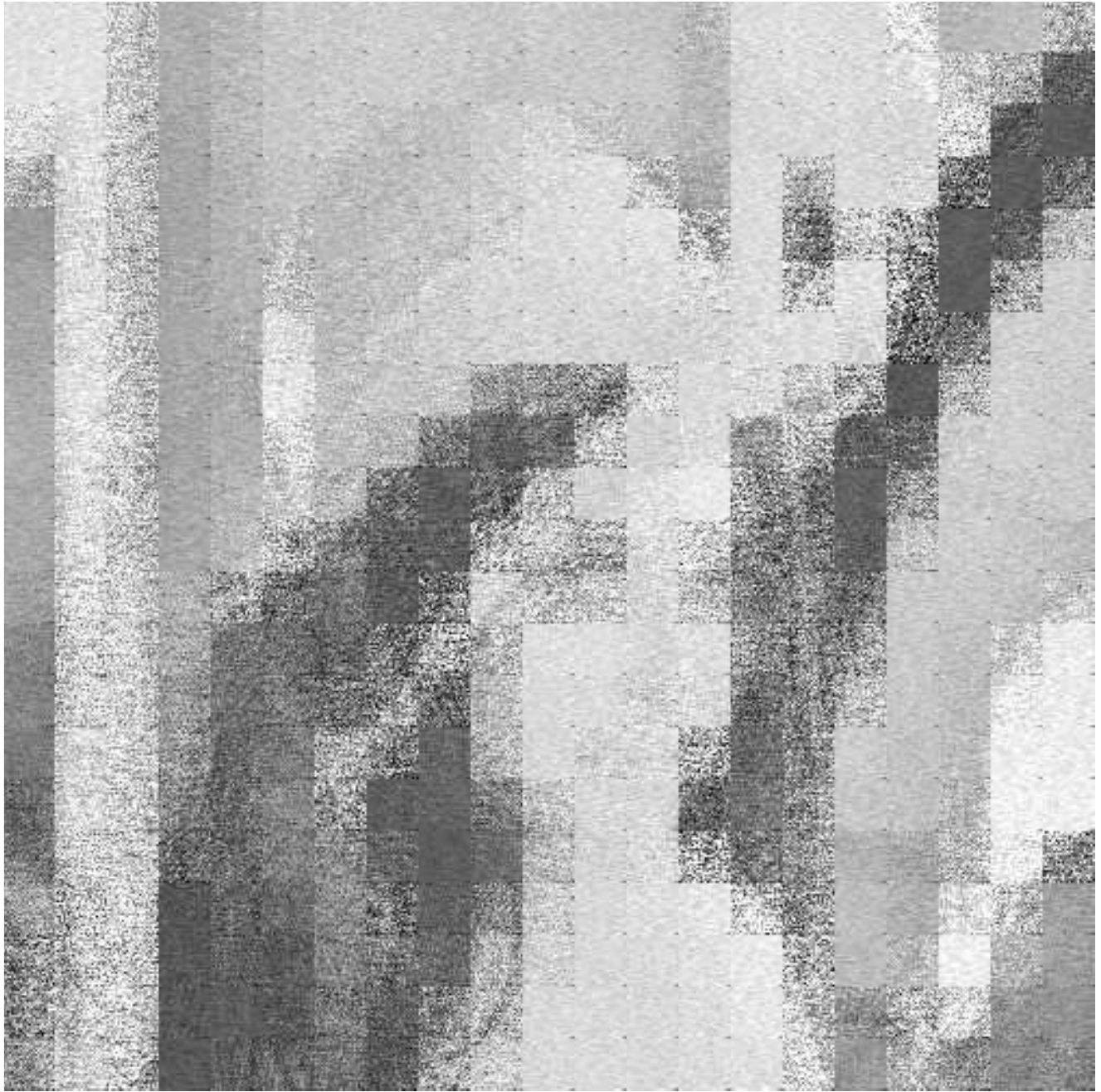}
  \includegraphics[scale=0.25]{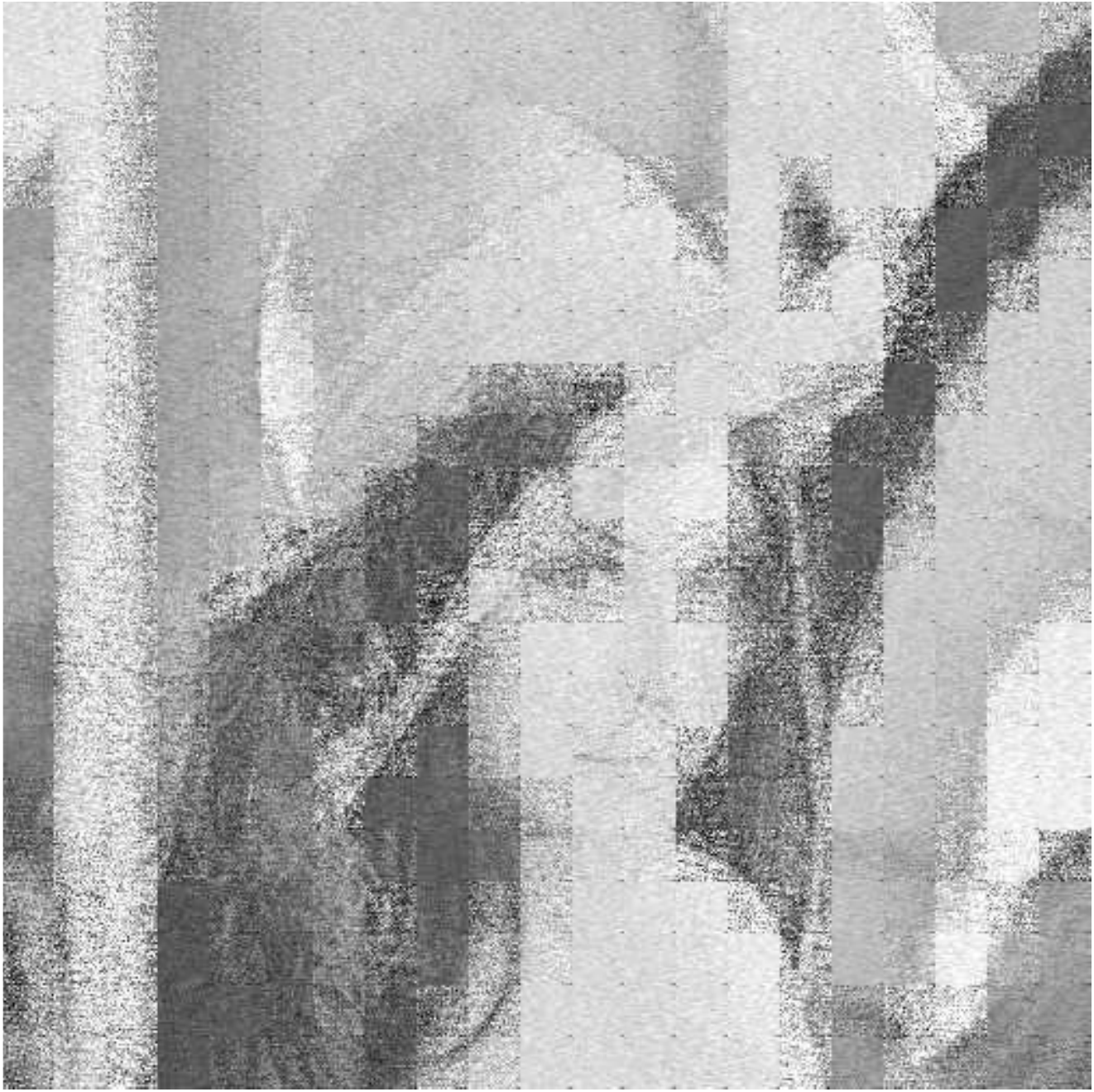}
  \includegraphics[scale=0.25]{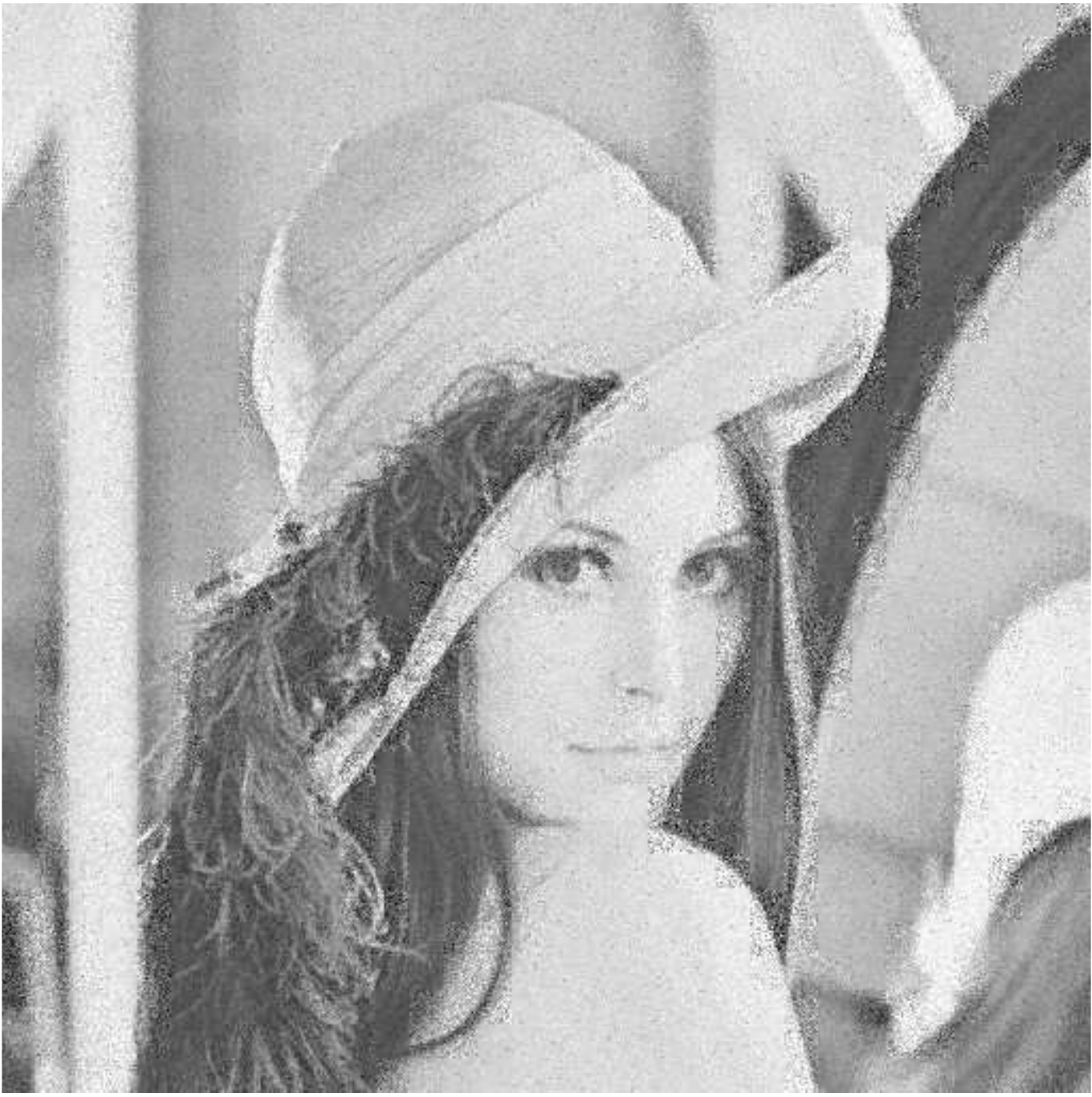}\\
 {\footnotesize (a) ~~~~~~~~~~~~~~~~~~~~~~~~~~~~~(b)~~~~~~~~~~~~~~~~~~~~~~~~~~~~~ (c)~~~~~~~~~~~~~~~~~~~~~~~~~~~~~(d)~~~~~~~~~~~~~~~~~~~~~~~~~~~~~(e) }\\
  \caption{{\footnotesize Cloud-assisted image encryption with different security level. (a) Original image of ``Lenna". (b) $n$-secure. (c)  $n/2$-secure. (d)  $n/3$-secure. (e) Recovery image of end user.}}\label{f_se_uniform}
  \end{center}
\end{figure*}

\subsubsection{Impact of Block Size}
In our experiment, the size of the division image block needs to consider the following three issues:

 1) The block size affects the secure performance of eCIS. The greater the size of division block is, the smaller the probability that the original signal is successfully recovered by the attacker.

 2) The block size  affects the scale of $\ell_1$ optimization problem, namely, Eq.\ref{e_cry_cs_dec}. The greater size  Eq.\ref{e_cry_cs_dec}  solves, the more time the problem requires.

 3) The block size affects the compression performance of image source and the communication cost between sampling device and cloud. The greater the size of division block is, the better the sparsity of compression block cloud obtain. The sparse level decides the number of transmission CS measurements, which affects the transmission cost from sampling device to cloud.


Fig.\ref{f_se_diff_block_size} displays the experiment results of our scheme with different block sizes and $n$-secure with uniform distribution random perturbation.
Fig.\ref{f_se_diff_block_size} (a1) and (a2) are the original image sources of ``Soccer" and ``Goldhill". Fig.\ref{f_se_diff_block_size} (b1) and (b2) display the cloud recovery image with $16\times 16$ pixels of block size. Fig.\ref{f_se_diff_block_size} (c1) and (c2) are  $24\times 24$ pixels of block size. Fig.\ref{f_se_diff_block_size} (d1) and (d2) are $32\times 32$ pixels of block size. Fig.\ref{f_se_diff_block_size} (e1) and (e2) are recovery image of end user.  According to Fig.\ref{f_se_diff_block_size}, it illustrates that the recovery images of cloud become more ambiguous with the increasing the block size.
Fig.\ref{f_psnr_diff_block_size} displays PSNR comparison of recovery image between cloud and end user across different block size. The experiment is implemented under the same number of measurements for the same block size. It shows that the recovery numerical performance of end user is much better than the cloud. In cloud, the PSNR of ``Soccer" and ``Goldhill" are less than 29dB and 29.5dB, respectively.

Fig.\ref{f_se_diff_1_block_size}  displays experimental result of cloud-assisted image encryption with $40\times 40$ pixels of block size of image ``soccer". The red image block of Fig.\ref{f_se_diff_1_block_size} (a) is our selected block. The experiment was implemented with $n$-secure and  uniform distribution random perturbation. The experimental results displays that the image details are invisible at all as shown in Fig.\ref{f_se_diff_1_block_size} (c).
However, each encryption block still represents average pixel values of the original image from visual aspect as shown in Fig.\ref{f_se_diff_block_size}. The reason is that the mainly energy of block under DCT basis focus on only one few low frequency coefficients. If the user wants to make the encryption image become more ambiguous from visual aspect, we could increase the division block size.

\begin{figure}[t]
  \begin{center}
  \includegraphics[scale=0.25]{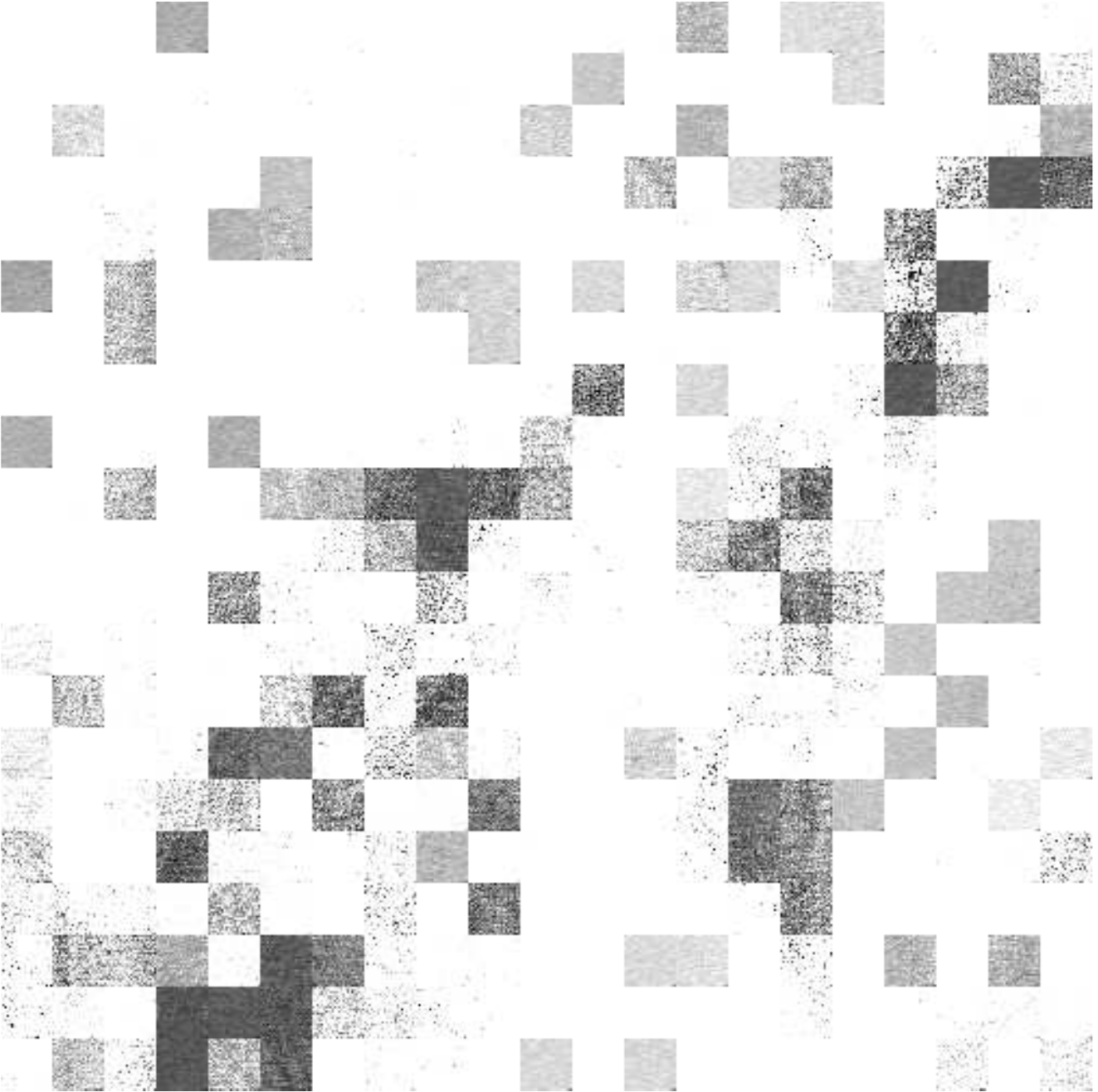}~~~~~~~~~
  \includegraphics[scale=0.25]{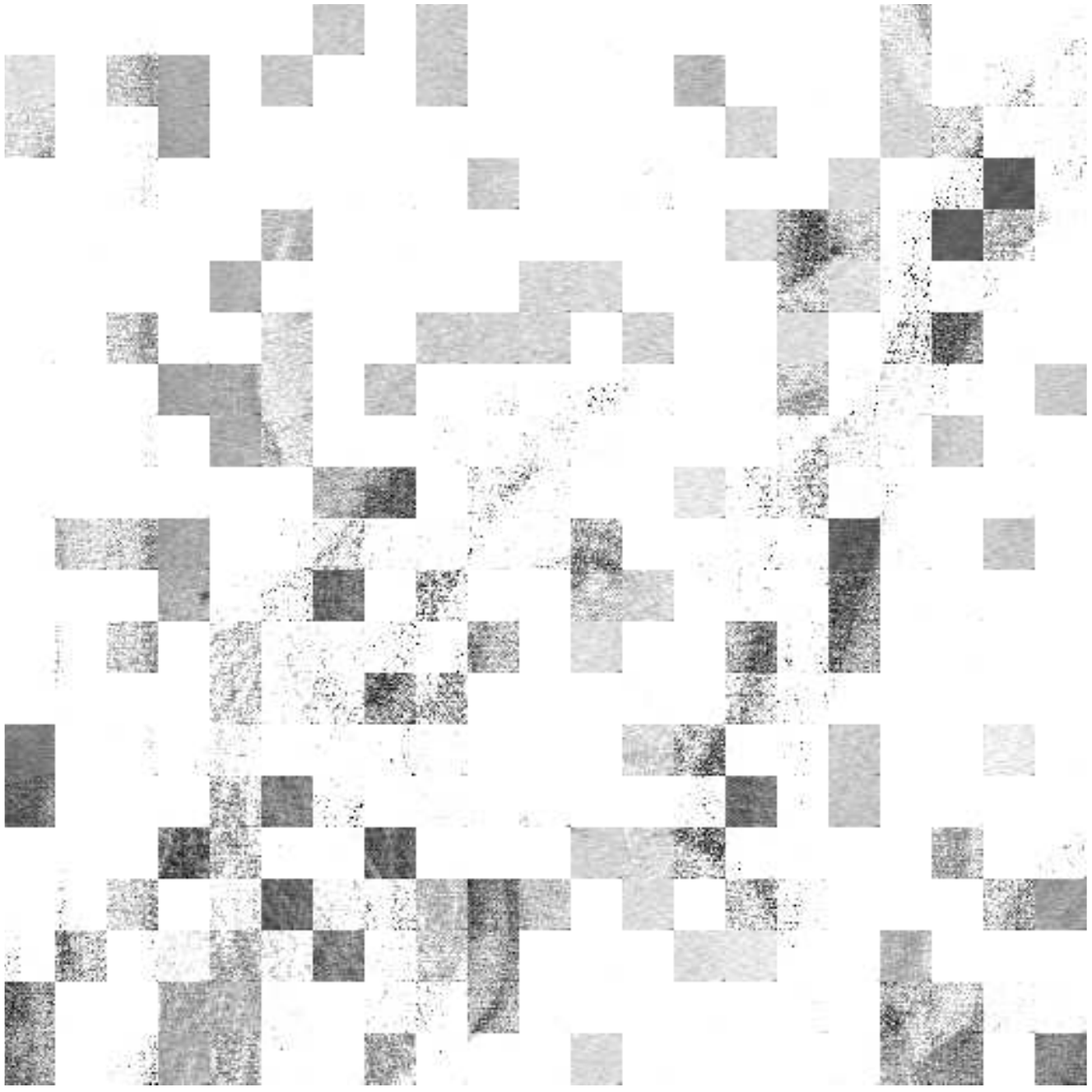}\\
{\footnotesize (a) ~~~~~~~~~~~~~~~~~~~~~~~~~~~~~~~~~~(b)}\\
   \end{center}
  \caption{{\footnotesize Cloud-assisted image encryption with uniform random perturbation and amplitude random encryption. (a) $n/2$-secure. b) $n/3$-secure.}}\label{f_se_alph_uniform}
\end{figure}

\subsubsection{Security Level Considerations}
In this part, we demonstrate  the experiment results of eCIS with different security levels. Fig.\ref{f_se_uniform} shows that
our results on  image ``Lenna" and under uniform distribution random perturbation. Fig.\ref{f_se_uniform} (a) displays the original image. Fig.\ref{f_se_uniform} (b), (c), and (d) display the recovered images using cloud sparse signal with $n$, $n/2$ and $n/3$-secure, respectively. Fig.\ref{f_se_uniform} (e) displays the recovered image at the end user. According to Fig.\ref{f_se_uniform}, the recovered  image  using cloud sparse signal leaks more details about the original image with increasing the security level. The attacker can  obtain almost contour and many details of original image when the encryption matrix is $n/3$-secure as shown in  Fig.\ref{f_se_uniform} (d).

In order to enhance the visual encryption effectiveness, we could further encrypt the amplitude of sparse signal with random multiplication. For example, we can adopt the encryption process as described in  Eq.\ref{f_en_rand_alpha} for each image block.
\begin{equation}\label{f_en_rand_alpha}
     A_k=\alpha\cdot\pi_k(I)
\end{equation}
where $\alpha$ is also a random value. It means that the encryption pixel values of each block are multiplied by a same random value. Fig.\ref{f_se_alph_uniform} shows the experiment results of ``Lenna" image with $n/2$ and $n/3$-secure. The value of $\alpha$ is random selected between $0$ and $1$. The experiment result displays that we only obtain a little image information from  visual aspect. However, it cannot improve much help from attack aspect because all the pixel values are carried out the same linear operation. The goal of random amplitude encryption only perturbs the image contour from visual aspect.
\begin{figure*}[t]
 \begin{center}
  \includegraphics[scale=0.235]{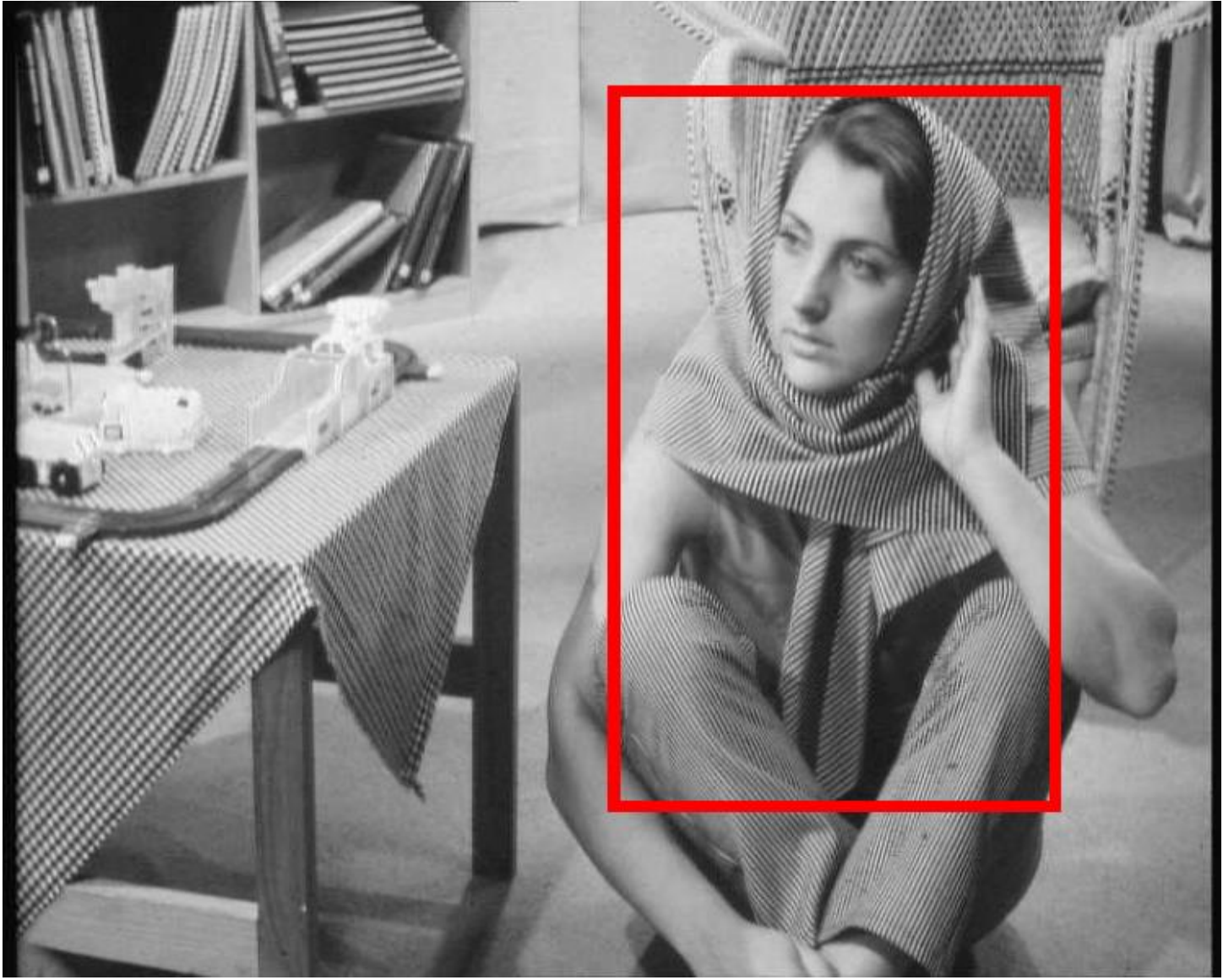}
  \includegraphics[scale=0.27]{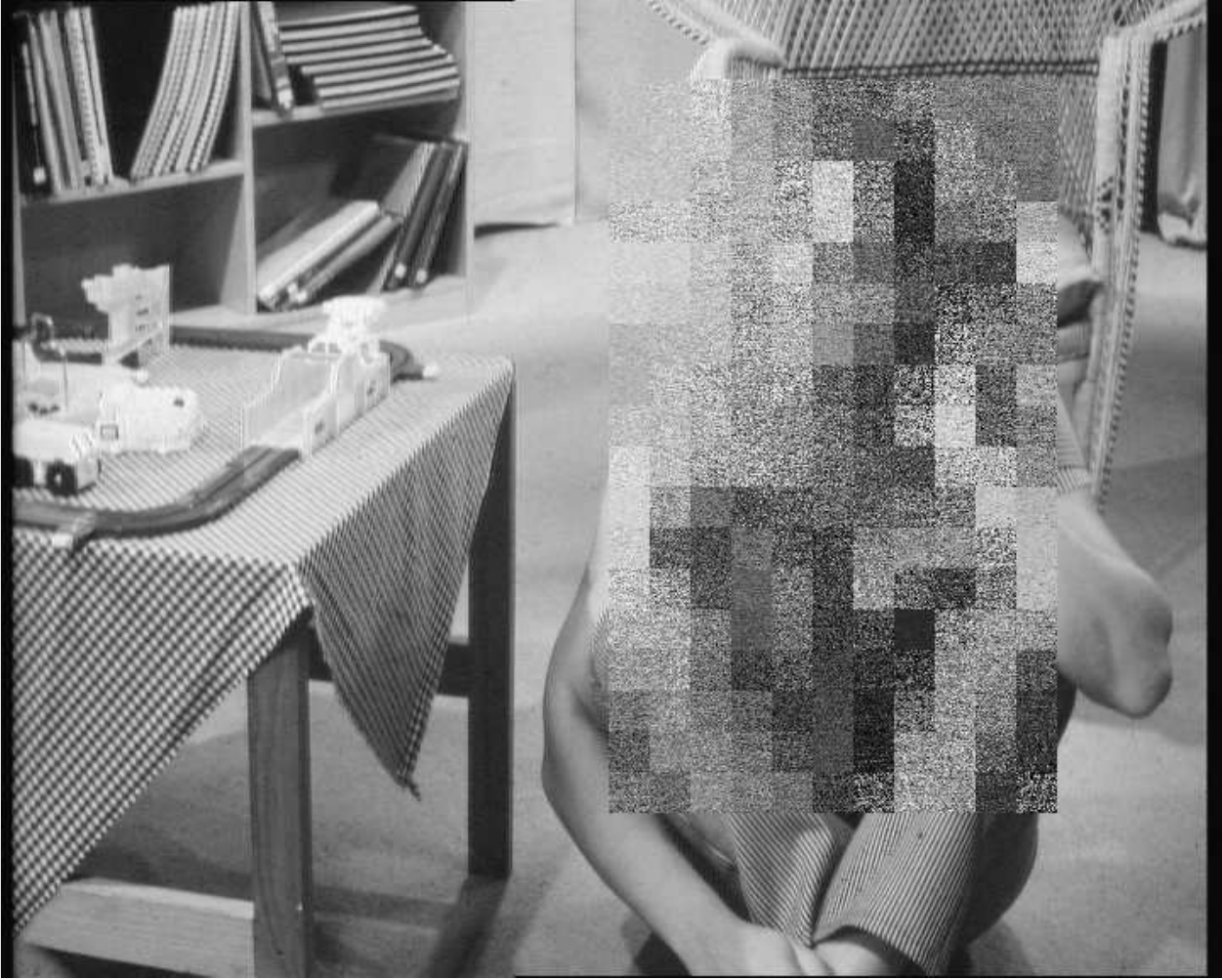}
  \includegraphics[scale=0.27]{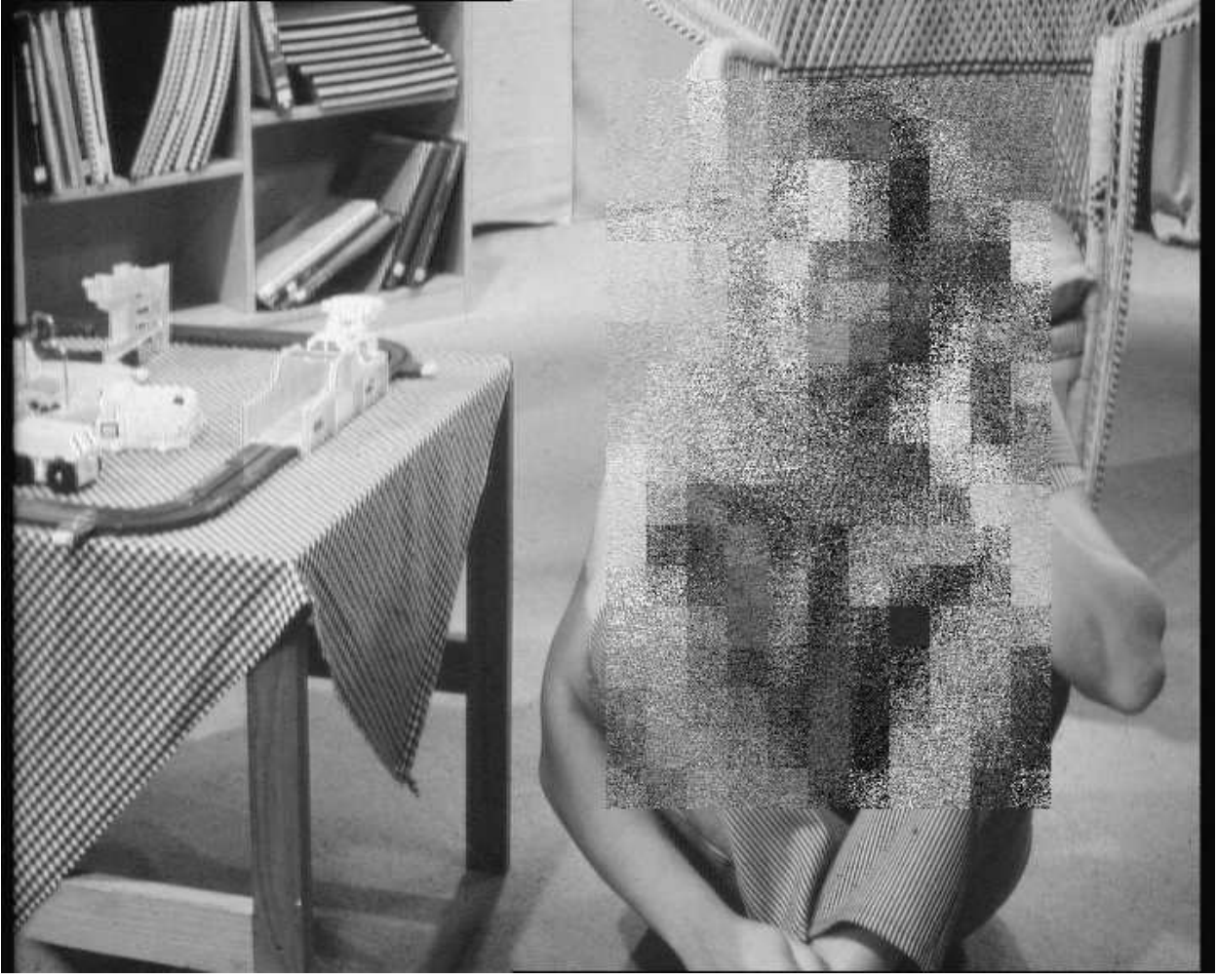}
  \includegraphics[scale=0.27]{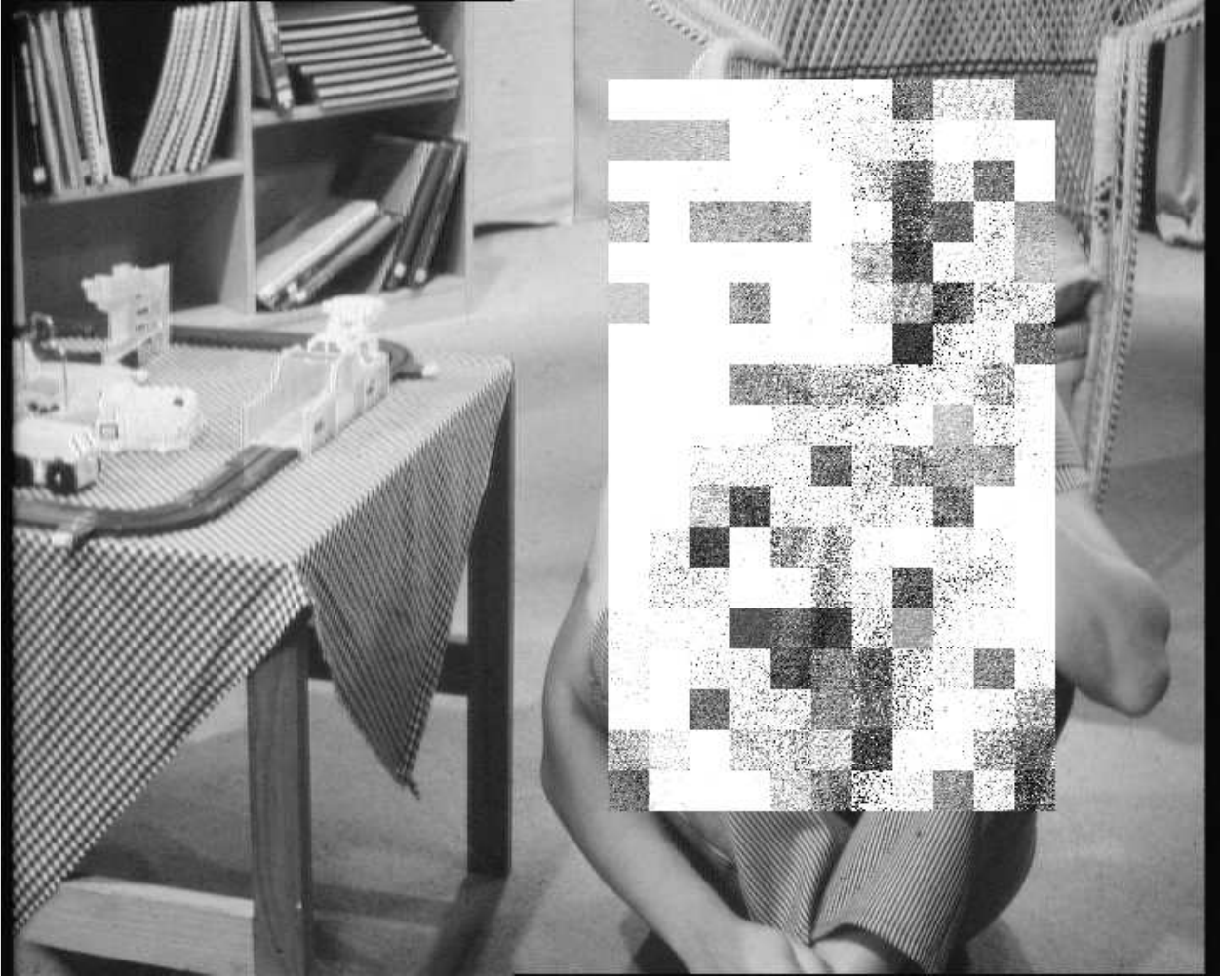}
  \includegraphics[scale=0.27]{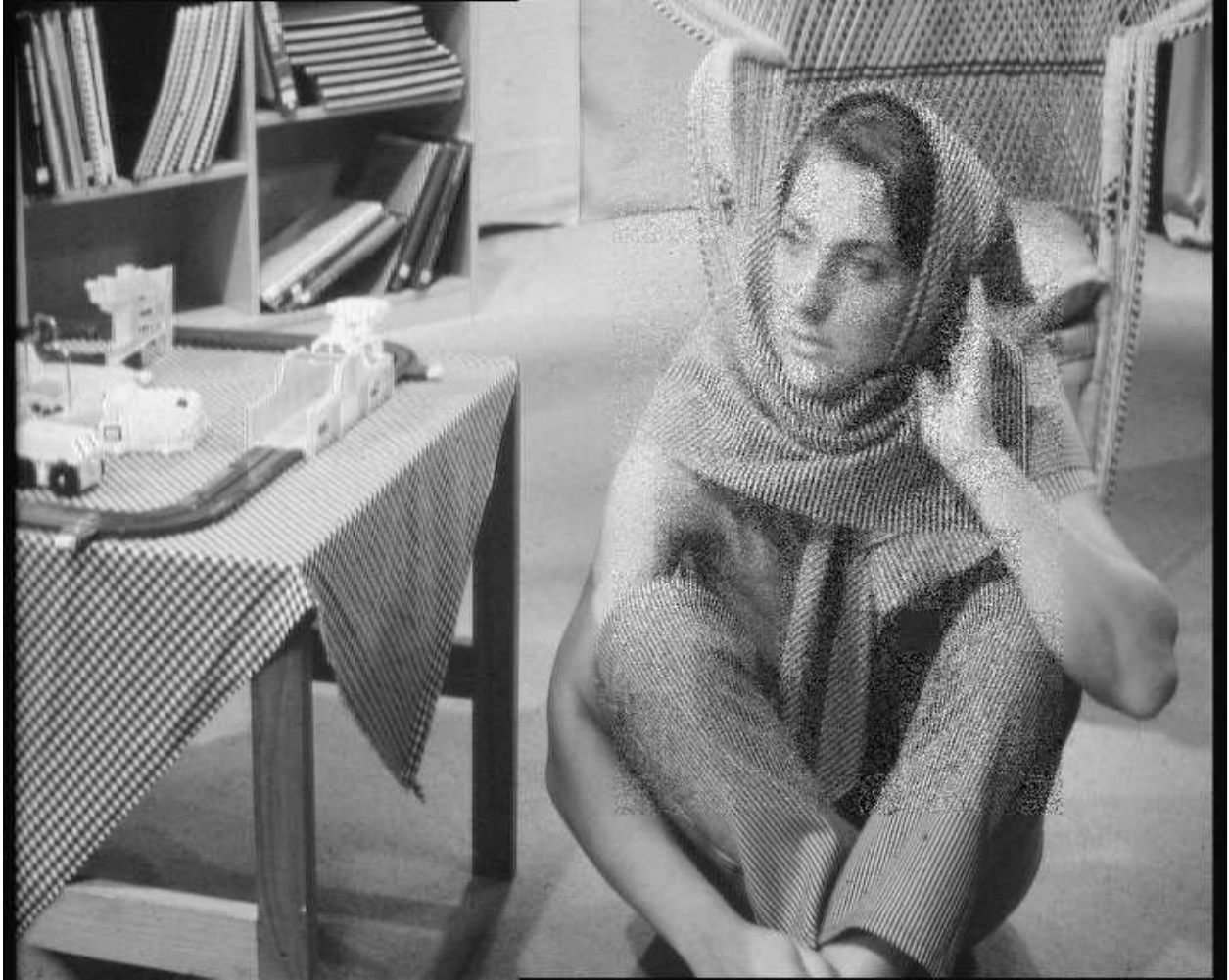}  \\
  {\footnotesize (a) ~~~~~~~~~~~~~~~~~~~~~~~~~~~~~(b)~~~~~~~~~~~~~~~~~~~~~~~~~~~~~ (c)~~~~~~~~~~~~~~~~~~~~~~~~~~~~~(d)~~~~~~~~~~~~~~~~~~~~~~~~~~~~~(e) }\\
  \caption{{\footnotesize Cloud-assisted image encryption with region of interest. (a) Original image and selected ROI of ``Barbara". (b) $n$-secure. (c) $n/2$-secure. (d) $n/2$-secure with random amplitude encryption. (e) Recovery image of end user.}}\label{f_se_ROI}
   \end{center}
\end{figure*}

\subsubsection{ROI Encryption}
eCIS can also  perform  encryption based on the user's  region of interest requirement. For example, Fig.\ref{f_se_ROI} displays the experiment result of our scheme with ROI image encryption and $24\times 24$ pixels of  block size. Fig.\ref{f_se_ROI} (a) is the original ``Barbara" image, the red block is the encryption block of ROI. Fig.\ref{f_se_ROI} (b), (c) and (d) are the recovered  images according to the cloud recovery sparse signal. Fig.\ref{f_se_ROI} (b) and (c) are $n$-secure and $n/2$-secure, respectively. Fig.\ref{f_se_ROI} (d) is the $n/2$-secure with random amplitude encryption and $0<\alpha<1$. Fig.\ref{f_se_ROI} (e) is the recovery image of end user. The experiment result displays that we almost can not know the details and contour of selected ROI according to Fig.\ref{f_se_ROI} (b) and (d). Fig.\ref{f_se_ROI} (d) is sufficient  ambiguous from visual aspect.

\begin{table*}
\begin{center}
\caption{Mean running time comparison of different block size (in seconds).}\label{t_run_time_comp}
\begin{tabular}[t]{|c|c|c|c|c|c|c|c|c|c|c|}
\hline
\multirow{2}{*}{Image} & \multirow{2}{*}{Block size} & \multicolumn{2}{c|}{Original\_CS} & \multicolumn{2}{c|}{Cloud\_Non\_encryption} & \multicolumn{2}{c|}{Our scheme} & \multicolumn{3}{c|}{Speedup}\tabularnewline
\cline{3-11}
 &  & $T_{sd}$ & $T_{eu}$ & ~~~$T_{sd}$ ~~ & $T_{eu}$ & $T_{sd}$ & $T_{rec}$ & $T_{sd}$ & $T_{eu}$ & $T_{total}$  \tabularnewline
\hline
\multirow{3}{*}{Lenna} & $24\times24$ & 0.0056  & 0.17824  & 0.00566  & 0.0263  & 0.00904  & 0.0344  & -1.7$\times$ & 5.2$\times$ & 4.2$\times$\tabularnewline
\cline{2-11}
 & $32\times32$ & 0.01828  & 0.64044  & 0.0168  & 0.0835  & 0.02278  & 0.13748  & -1.3$\times$ & 4.6$\times$ & 4.1$\times$\tabularnewline
\cline{2-11}
 & $48\times48$ & 0.08816  & 5.73816  & 0.0884  & 0.40624  & 0.1125  & 0.79696  & -1.3$\times$ & 7.2$\times$ & 6.4$\times$\tabularnewline
\hline
\multirow{3}{*}{Soccer} & $24\times24$ & 0.0056 & 0.16688 & 0.00498  & 0.02654  & 0.00694  & 0.03284  & -1.2$\times$ & 5.1$\times$ & 4.3$\times$\tabularnewline
\cline{2-11}
 & $32\times32$ & 0.0175  & 0.65472  & 0.01784  & 0.08304  & 0.02808  & 0.14042  & -1.6$\times$ & 4.7$\times$ & 4.0$\times$\tabularnewline
\cline{2-11}
 & $48\times48$ & 0.08764  & 5.8582  & 0.09118  & 0.41128  & 0.11528  & 0.76554  & -1.3$\times$ & 7.7$\times$ & 6.8$\times$\tabularnewline
\hline
\end{tabular}
\end{center}
\end{table*}

\subsection{Overhead Evaluation}
In this subsection, we evaluate the overhead of eCIS. We mainly focus on computation cost at the sampling device and end user.  In order to effectively evaluate the overhead of our scheme, we compare eCIS with the following existing schemes:
(1) CS-based image process without cloud service and  security consideration (Orignial\_CS),
(2) Cloud-assisted and CS-based image process without security consideration (Cloud\_Non\_encryption),
and (3) Cloud-assisted and CS-based image process with security consideration by LP problem transformation  \cite{wang2014privacy}.

For ease of presentation, we implement different schemes with the same size of image block, $24\times24$, $32\times32$, and $48\times48$ pixels of block size. All the results represent the mean of $50$ experiments and each experiment is randomly selected one block from the same test image. We implement our scheme with uniformly random perturbation and $n$-secure. OMP algorithm is selected to solve $\ell_1$ optimization problem \cite{OMP}. Let $t_{sd}$ and $t_{eu}$ denote the running time of sampling device and  end user, respectively. In this experiment, we do not consider the image sampling time. $t_{sd}$ is the image compression time for Original\_CS and Cloud\_Non\_encryption schemes. For our scheme, $t_{sd}$ is the compression and encryption time. $t_{eu}$ is the time of solving $\ell_1$ optimization problem and image recovery for Original\_CS scheme. $t_{eu}$ is only the time of image recovery for Cloud\_Non\_encryption schemes.  In our scheme, $t_{eu}$ is the time of image decryption and recovery.

Table.\ref{t_run_time_comp} displays the mean running time (in seconds) comparisons of Orignial\_CS, Cloud\_Non\_encryption, and eCIS. The last column of Table.\ref{t_run_time_comp} represents the system speedup compared our scheme with Orignial\_CS scheme. The first and the second sub-columns represent the time speedup of sampling device and end user, respectively. The last sub-column is the total time ($T_{total}=T_{sd}+T_{eu}$) speedup without considering the time spent on the  cloud. According to Table.\ref{t_run_time_comp}, our running time  only increases  1.2$\times\sim1.7\times$  to obtain encryption function for sampling device under different block sizes compared with Orignial\_CS scheme.  In \cite{wang2014privacy}, LP problem transformation scheme requires 5.6$\times$ and 9.4$\times$ time cost compared with Orignial\_CS scheme to obtain encryption function for $32\times 32$, $48\times 48$ pixels of block sizes, respectively. Meanwhile, our scheme can decrease $4.6\times\sim7.7\times$ time cost for end user compared with Orignial\_CS scheme. For total time speedup, the experimental result displays that our eCIS  decrease $4.1\times\sim6.8\times$ time cost.  Even when compared  with Cloud\_Non\_encryption schemes, our running time only increases by up to $2\times$ in end user. In \cite{wang2014privacy}, the authors didn't consider the final image recovery cost. They decreases the total running time $4.0\times$ and $3.4\times$ with $32\times 32$ and $48\times 48$ pixels of block size, respectively. Without considering  the image recovery time, our scheme decreases the total running time $8.9\times$ and $11.7\times$ with $32\times 32$ and $48\times 48$ pixels of block size of ``Lenna", respectively. For the image of ``Soccer", our scheme decrease the total running time by $8.37\times$ and $12.7\times$ with $32\times 32$ and $48\times 48$ pixels of block sizes, respectively. The experiment result shows that eCIS can keep the user's image privacy with low-complexity  compared with the existing cloud-assisted image service scheme.

\section{Related Work}
\textbf{Image compression and encryption:}
Image compression technology can be divided into two categories, transform-based compression and CS-based compression. Existing image encryption technology is mainly aimed at transform-based compression image such as \cite{chen2010optical,liu2011image,wang2011new}. Transform-based compression technology requires high computation complexity and storage cost for encoder. This type of image compression technology is not suitable for resource-constrained smart device. CS-based image compression technology can shift the complexity from encoder to decoder. Although  CS theory and its applications have received lots of researches in recent years \cite{candes2009lr,needell2009cosamp,chen2014robust,zhang2009spatio,wxb2012energy}, CS-based image encryption technology also require high computation complexity \cite{huang2011robust,zhang2011compressing}. These methods are not  appropriate for resource-constrained smart device.
For resource-constrained smart device user, these two types of compression and encryption technology  not  directly meet the  requirement of user's privacy and system resource.

\textbf{Secure computation outsourcing:}
With the development of cloud computing in recent years, cloud provides an new avenue for storage and  computation outsourcing according to the user's demand \cite{armbrust2010view}. One advantage of cloud paradigm is resource outsourcing for resource-constrained smart device. However, cloud is public and exists lots of security threat such as data transmission security, data storage security and so on \cite{zhang2010cloud,subashini2011survey}. Cloud-based computation outsourcing has also been studied by many researchers. For example,  C. Wang  \emph{et. al} have proposed a secure and practical outsourcing of linear programming scheme in cloud computing \cite{wang2011secure}. In \cite{gennaro2010non,atallah2010securely}, the authors also proposed many schemes to encrypt input and output and implement secure computation outsourcing. These methods are all following the thought of fully homomorphic encryption \cite{gentry2009fully}, and transform the original LP problem into another LP problem. Since CS decoding is equivalent to solve LP problem, C. Wang \emph{et.al.} proposed a cloud-assisted computation outsourcing scheme for healthcare video monitoring \cite{wang2014privacy}. This proposal also require high complexity transformation operation to implement secure computation outscourcing.

\section{Conclusion and Future work}
In this paper, we discussed a novel  encrypressive low-complexity cloud-assisted image service scheme via compressive sensing. Although traditional image encryption  and secure linear programming outsourcing techniques can efficiently protect image privacy, these methods are not appropriate for resource-constrained device because of high computational cost.  Our scheme can efficiently transform the computation and storage cost to the cloud without increasing transmission cost, and protect image privacy according to user's adaptive security demand.  Extensive experiment results demonstrated our scheme can significantly save the system running time. In our scheme, we only consider one-dimension CS-based image compression and encryption. Future work will extend to two-dimension image compression and encryption to further reduce network transmission cost.

\end{document}